\documentclass[lettersize,journal]{IEEEtran}
\usepackage{amsmath,amsfonts}
\usepackage{algorithmic}
\usepackage{algorithm}
\usepackage{array}
\usepackage[caption=false,font=normalsize,labelfont=sf,textfont=sf]{subfig}

\usepackage{textcomp}
\usepackage{stfloats}
\usepackage{url}
\usepackage{verbatim}
\usepackage{hyperref}
\usepackage{graphicx}
\usepackage{cite}
\usepackage{booktabs} 
\begin{document}

\title{From Provable to Practical: A Problem-Driven Survey of Classical and Machine-Learning Defenses for DV/CV Quantum Key Distribution}

\author{Hasan~Abbas~Al-Mohammed and Afnan~S.~Al-Ali%
\thanks{H.~A.~Al-Mohammed and A.~S.~Al-Ali are with  Qatar University, Doha, Qatar (e-mail: ha1800217@qu.edu.qa; afnan.alali@qu.edu.qa).}%
}


\maketitle

\begin{abstract}
Quantum key distribution (QKD) promises information-theoretic security, yet practical deployments in both discrete-variable (DV) and continuous-variable (CV) settings remain exposed to device imperfections, channel manipulation, finite-key effects, and the brittleness of machine-learning (ML) components introduced for adaptation and monitoring. This survey adopts a strictly \emph{problem-driven} perspective: nine practical problem classes (P1–P9) spanning five layers—device/measurement, channel/physical, protocol/finite-key, ML, and network/coexistence—are each analyzed under a uniform \emph{problem $\rightarrow$ classical solution $\rightarrow$ ML solution $\rightarrow$ comparative analysis $\rightarrow$ gaps} template. Within this template we contrast transparent physical mitigations (optical filtering, power monitoring, calibration hardening, MDI/DI, impairment-aware planning) with ML-enabled adaptation and screening (drift-robust parameter prediction, anomaly and attack detection, noise prediction and channel/core allocation, adversarial purification). Where the corpus reports quantitative evidence, we surface it directly: e.g., DBSCAN-based CV attack detection at $\mathrm{P}{=}99.7\%/\mathrm{R}{=}99.8\%/\mathrm{F1}{=}0.998$; APE-GAN/KRMNMF restoring robust accuracy to $74.88\%/71.6\%$ (combo $78.8$–$79.5\%$) under adversarial perturbations; tree-based channel-amplification detection at $100\%/91.26\%$ in low/high-noise regimes with post-selection gains of $+0.024$ bits/pulse ($\sim 6$ kbps at 100~MHz); and LightGBM-style noise predictors reducing evaluation time by up to $98.8\%$ with $\le 25\%$ error against actual lowest-noise channels. From this evidence we (i) build a master comparison table covering P1–P9 plus free-space stabilization (V-C), CV defect identification (P1-H), CV metaheuristic+ML (P4-M), and quantum-native ML for DV (P4-Q); (ii) propose a TETCI-friendly benchmarking grid coupling datasets (fiber/free-space; lab/field), stress protocols (adversarial, distribution-shift, small-block finite-key), and a unified metric stack (P/R/F1, $\Delta$SKR and maximum distance at matched operating points, latency/compute, robustness); and (iii) translate the synthesis into actionable defense-in-depth guidelines that orthogonally combine classical and ML layers, couple detector decisions to finite-key bounds, and constrain ML inference within controller cycles. We close by mapping near-, mid-, and long-term research priorities—public telemetry and adversarial suites, hardware-in-the-loop evaluation, head-to-head MDI/DI vs.\ hardened-link finite-key studies, and privacy-preserving federated learning—as preconditions for auditable, real-world QKD deployments.
\end{abstract}


\section{Introduction}
\IEEEPARstart{Q}{uantum} key distribution (QKD) is the most operationally mature branch of quantum cryptography, offering—under idealized models—information-theoretic security rooted in the laws of quantum mechanics rather than in computational hardness assumptions. Two principal modalities, \emph{discrete-variable (DV)} and \emph{continuous-variable (CV)} QKD, have matured through more than two decades of foundational theory, security proofs, and increasingly capable implementations \cite{num04_Gisin2002_QuantumCryptographyReview, num29_Li2017_CVQKD_ChinesePhysicsB_Review, num96_Scarani2009_Security_PracticalQKD_RMP,num13_Xu2020_SecureQKDRealisticDevices}. Yet a persistent gap separates \emph{provable} security—derived under idealized device and channel models—from the \emph{practical} security of real fielded systems, where finite-precision detectors, calibration drift, free-space turbulence, coexisting classical traffic, and adversarial environments all push the operating envelope away from the proof's assumptions. This gap does not narrow automatically as the technology scales; it widens as deployments diversify from controlled fiber testbeds toward metropolitan backbones, satellite links, multi-core and multi-band coexistence with classical optics, and resource-constrained IoT and railway scenarios. A \emph{problem-centric} view, grounded in measured device and network behavior rather than in abstract security models alone, is therefore essential.

In practice, this gap has enabled a now-substantial catalog of \emph{quantum hacking} strategies that exploit imperfections in sources, modulators, detectors, and the optical channel rather than violating the underlying quantum-mechanical assumptions. The two modalities exhibit distinct but structurally analogous attack surfaces. For \textbf{CV-QKD}, the local oscillator (LO) is both an enabler of high-rate coherent detection and a primary vulnerability: documented attacks include LO fluctuation, wavelength manipulation, calibration tampering, detector saturation, and homodyne-detector blinding, each capable of biasing parameter estimation in directions that conceal eavesdropping while still satisfying nominal acceptance thresholds~\cite{num05_Ma2013_LOFluctuationLoophole_CVQKD,num06_Huang2013_WavelengthAttack_CVQKD,num07_Ma2013_WavelengthAttack_Heterodyne_CVQKD, num08_Jouguet2013_PreventingCalibrationAttacks_LO_CVQKD,num09_Qin2016_SaturationAttack_CVQKD,num10_Qin2018_HomodyneBlinding_CVQKD}. For \textbf{DV-QKD}, the analogous pressure point is the single-photon detector, with bright-illumination, after-gate, time-shift, and phase-remapping attacks exploiting timing and saturation behavior to convert a quantum measurement into a partially classical, controllable channel \cite{num14_Lydersen2010_HackingCommercialQKD_BrightIllumination,num48_Wiechers2011_AfterGateAttack_NJP,num50_Qi2007_TimeShiftAttack_QIC,num97_Fung2007_PhaseRemappingAttack_PRA,wang2025advances}. Running in parallel with this offense, CV-QKD demonstrations have steadily expanded the reach of the technology—long-distance operation with excess-noise control, real-environment monitoring, and local local-oscillator (LLO) operation over tens of kilometers of fiber—\cite{num11_Huang2016_LongDistanceCVQKD_ExcessNoiseControl,num12_Jouguet2013_LongDistanceCVQKD_ExperimentalDemo,num72_Liu2017_Monitoring_CVQKD_RealEnvironment,num02_Hajomer2022_CVQKD60kmRLO,zhu2025remote}. The implication for defense evaluation is sharp: countermeasures must be judged not only by how well they \emph{detect} an attack but by how much \emph{secret-key rate (SKR)}, distance, and finite-key headroom they cost to deploy, since a defense that restores robustness only by halving the SKR or doubling the controller latency is rarely deployable in practice.

Two lines of defense have evolved against this attack catalog, with very different epistemic commitments. \textit{Classical} countermeasures act at the physical or protocol layer and are, in the most desirable cases, \emph{transparent}—their effect on the security model is auditable and bounded. They range from improved calibration and excess-noise control to fundamental architectural shifts: hardened LO handling and excess-noise budgeting~\cite{num11_Huang2016_LongDistanceCVQKD_ExcessNoiseControl,num08_Jouguet2013_PreventingCalibrationAttacks_LO_CVQKD}, and the more ambitious \emph{device-independent (DI)} and \emph{measurement-device-independent (MDI)} paradigms that relocate trust away from vulnerable modules at the cost of additional complexity~\cite{num16_Acin2007_DeviceIndependentSecurity_CollectiveAttacks,num17_Lo2012_MDIQKD_PRL,num28_Tang2014_MDIQKD_Polarization_Experimental}. In parallel, a fast-growing body of work applies \textit{machine learning (ML)} to configuration, monitoring, and active defense, motivated by the observation that many practical impairments are nonstationary, multidimensional, and only weakly captured by closed-form models. The reported gains span parameter prediction and real-time optimization~\cite{num21_Liu2018_ML_AutomaticParameterPrediction_CVQKD,num23_Wang2019_ML_OptimalParameterPrediction_QKD,num24_Ding2020_RandomForest_ParamPrediction_QKD, num25_Su2019_BP_NN_ParameterOptimization_Atmosphere_CVQKD}, stabilization and field monitoring~\cite{num72_Liu2017_Monitoring_CVQKD_RealEnvironment,num41_Liu2019_PhaseModulationStabilization_ML_QKD}, attack detection and multi-attack classification~\cite{num30_Mao2020_ML_DetectingQuantumAttacks_CVQKD,num31_Du2022_MultiAttackDetection_NN_CVQKD, num37_Liao2022_DBSCAN_DetectingAttacks_CVQKD}, adversarial-perturbation defenses tailored to CV-QKD~\cite{num32_Tang2023_APE_GAN_Defense_CVQKD, num58_Fu2023_KRMNMF_Defense_CVQKD}, ML-assisted sifting on short samples~\cite{num98_Wu2022_SiftingScheme_CVQKD_ShortSamples_JOSAB}, and noise forecasting for QKD coexistence in dense optical networks~\cite{num78_Wang2022_NoisePrediction_ML_Secured_SWDM_B5G_Fronthaul,num81_Niu2019_LightGBM_NoiseSuppressing_DWDM_QKD}. Recent contributions further consolidate the role of ML and AI in QKD optimization, attack detection, robust operation, and quantum-safe integration~\cite{al2026machine,mohamed2026optiqkd,gupta2026ai}. ML, however, is not a free upgrade: it imports a new attack surface (distribution shift, adversarial examples, training-data poisoning) and a new accountability burden (auditability, reproducibility, and certifiability of learned components) that pure physical countermeasures do not share. The interesting question for this survey is therefore not \emph{classical vs.\ ML}, but \emph{when each is sufficient on its own, when they must be combined, and how to evaluate the combination against the same yardstick of SKR, distance, latency, and robustness}.

This survey focuses on \emph{practical security} for DV- and CV-QKD with an emphasis on the unified pipeline \emph{problem $\rightarrow$ classical solution $\rightarrow$ ML solution $\rightarrow$ comparative analysis $\rightarrow$ gaps}. Compared with existing QKD security reviews, which tend either to enumerate quantum-hacking attacks in isolation or to catalog ML applications without anchoring them to security-impact metrics, our contribution is to bind these two literatures into a single accounting framework whose unit of comparison is operational: detection quality \emph{and} the secret-key cost of achieving it. Specifically, we:
\begin{itemize}
    \item \textbf{Taxonomize practical problems} into five layers—device/measurement side-channels, channel/physical-layer manipulation, protocol/process vulnerabilities, ML-layer threats, and network/integration issues—with concrete DV and CV exemplars and a threat–mitigation map that contrasts the two modalities at each layer~\cite{num05_Ma2013_LOFluctuationLoophole_CVQKD,num06_Huang2013_WavelengthAttack_CVQKD,num07_Ma2013_WavelengthAttack_Heterodyne_CVQKD, num08_Jouguet2013_PreventingCalibrationAttacks_LO_CVQKD,num09_Qin2016_SaturationAttack_CVQKD,num10_Qin2018_HomodyneBlinding_CVQKD,num14_Lydersen2010_HackingCommercialQKD_BrightIllumination,num48_Wiechers2011_AfterGateAttack_NJP,num50_Qi2007_TimeShiftAttack_QIC,num97_Fung2007_PhaseRemappingAttack_PRA}.
    \item \textbf{Synthesize classical defenses} and quantify, where the corpus permits, their costs in SKR, distance, calibration effort, and architectural complexity, rather than restating them as a feature list~\cite{num11_Huang2016_LongDistanceCVQKD_ExcessNoiseControl,num08_Jouguet2013_PreventingCalibrationAttacks_LO_CVQKD,num16_Acin2007_DeviceIndependentSecurity_CollectiveAttacks,num17_Lo2012_MDIQKD_PRL,num28_Tang2014_MDIQKD_Polarization_Experimental}.
    \item \textbf{Review ML-enabled approaches} for parameter prediction, stabilization and monitoring, attack and defect detection, and adversarial-perturbation defense in CV—and, where evidence exists, DV—settings, surfacing the reported numerical performance so that head-to-head comparisons across the corpus become possible~\cite{num72_Liu2017_Monitoring_CVQKD_RealEnvironment,num21_Liu2018_ML_AutomaticParameterPrediction_CVQKD,num23_Wang2019_ML_OptimalParameterPrediction_QKD,num24_Ding2020_RandomForest_ParamPrediction_QKD, num25_Su2019_BP_NN_ParameterOptimization_Atmosphere_CVQKD,num41_Liu2019_PhaseModulationStabilization_ML_QKD,num30_Mao2020_ML_DetectingQuantumAttacks_CVQKD,num31_Du2022_MultiAttackDetection_NN_CVQKD, num37_Liao2022_DBSCAN_DetectingAttacks_CVQKD,num32_Tang2023_APE_GAN_Defense_CVQKD, num58_Fu2023_KRMNMF_Defense_CVQKD,num98_Wu2022_SiftingScheme_CVQKD_ShortSamples_JOSAB}.
    \item \textbf{Distill evaluation and deployment} requirements, with attention to data quantity and quality, robustness under distribution shift and adversarial pressure, controller-cycle latency budgets, and the integration of QKD-layer alerts with classical intrusion detection and routing at the network layer~\cite{num99_Alahmadi2022_CyberSecurity_Threats_SideChannel_DigitalAgriculture_Sensors,num100_Chkirbene2020_TIDCS_DynamicIDS_FeatureSelection_Access,num101_Hijazi2018_DL_IDS_IndustryNetwork_BDCSIntell,num102_Vacca2012_ComputerInformationSecurityHandbook,num103_Liao2013_IDS_ComprehensiveReview_JNCA,num104_Tama2019_TSEIDS_TwoStageClassifier_AnomalyIDS_Access,num105_GarciaTeodoro2009_AnomalyBased_IDS_Survey_CompSec,num106_Gumusbas2021_Survey_Databases_DL_Cybersecurity_IDS_IEEE_SystemsJ,num109_Sirisha2023_EfficientMLTechniques_IDS_ICSCDS,num110_Srilakshmi2022_SecureOptimizationRouting_MANETs_Access,num111_Hung2020_EnergyEfficient_CooperativeRouting_Heterogeneous_WSN_Access}.
\end{itemize}

\noindent\textbf{Paper organization.}
Section~\ref{sec:background-threat} reviews DV/CV background and the threat model.
Section~\ref{sec:taxonomy} defines the taxonomy of practical problems.
Section~\ref{sec:methodology} details the survey methodology.
Section~\ref{sec:problem-centric} forms the core: each subsection follows a Problem, Classical, ML, and Gaps structure for specific attacks and operational issues (DV and CV).
Section~\ref{sec:benchmarking} proposes a benchmarking and evaluation framework.
Section~\ref{sec:systems} distills systems-integration lessons (coexistence, monitoring, IDS).
Section~\ref{sec:guidelines} offers design guidelines.
Section~\ref{sec:open} highlights open challenges.
Section~\ref{sec:conclusion} concludes.


\section{QKD Background \& Threat Model}
\label{sec:background-threat}

\subsection{QKD primer: DV vs.\ CV, MDI/DI, and finite-key}
QKD comprises two principal modalities—DV and CV—each with its own trade-off geometry across security assumptions, hardware complexity, and reachable distance/rate, as documented in a broad body of theory and implementation work \cite{num04_Gisin2002_QuantumCryptographyReview, num29_Li2017_CVQKD_ChinesePhysicsB_Review, num13_Xu2020_SecureQKDRealisticDevices, num96_Scarani2009_Security_PracticalQKD_RMP,wang2025quantum,wang2024high}. Within CV-QKD, the architectural choice between transmitted local oscillator (TLO) and local local oscillator (LLO) is consequential beyond engineering convenience: TLO designs simplify phase reference recovery but expose a strong classical channel co-propagating with the quantum signal, while LLO designs eliminate that channel at the price of more demanding phase locking. The LLO direction has now been pushed to nontrivial fiber distances—e.g., an LLO-based CV-QKD demonstration over 60~km with measurable secret-key rate~\cite{num02_Hajomer2022_CVQKD60kmRLO}—narrowing the practicality gap with TLO while reducing the LO-side attack surface. Architectural hardening via \emph{device-independent (DI)} and \emph{measurement-device-independent (MDI)} paradigms goes further still, relaxing trust in specific modules (typically the measurement device) at the cost of additional resources and, for DI, very stringent loophole-free Bell-test requirements~\cite{num16_Acin2007_DeviceIndependentSecurity_CollectiveAttacks,num17_Lo2012_MDIQKD_PRL,num28_Tang2014_MDIQKD_Polarization_Experimental}. The reader should view DV/CV $\rightarrow$ TLO/LLO $\rightarrow$ MDI/DI as a \emph{trust gradient}: each step moves assumptions further away from vulnerable physical modules, but each step also costs SKR, complexity, or both, and these costs interact with the \emph{finite-key} penalty that any real-block-length system must pay~\cite{num74_Furrer2012_FiniteKey_ComposableSecurity_CoherentAttacks_PRL,num75_Leverrier2017_Security_CVQKD_GaussianDeFinetti}. System-level analyses make this concrete \cite{num96_Scarani2009_Security_PracticalQKD_RMP,num13_Xu2020_SecureQKDRealisticDevices}, while application-oriented QKD protocol studies help bridge the theoretical security model and real-life deployment requirements, particularly for authentication, key management, and network integration~\cite{al2024qkd}.

\begin{figure}[!t]
\centering
\includegraphics[width=3.5in]{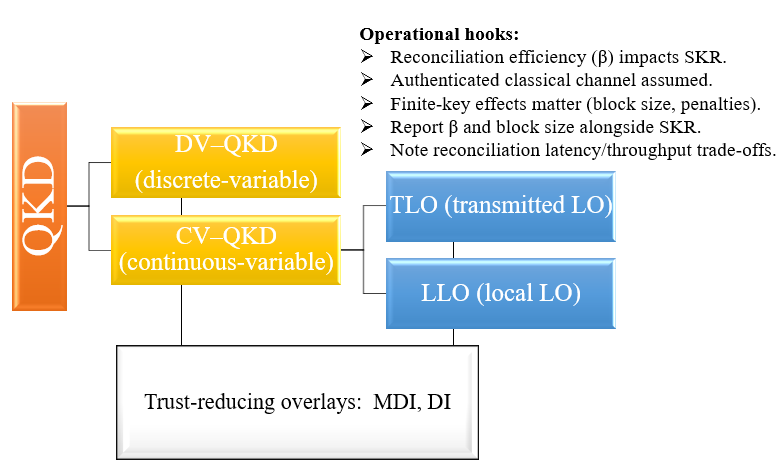}
  \caption{QKD modalities and trust assumptions.}
  \label{fig:modality-trust}
\end{figure}
Reconciliation and authentication are integral to practical operation: reconciliation efficiency $\beta$ directly impacts achievable SKR, and authenticated classical exchanges are assumed throughout~\cite{num03_Leverrier2008_MultidimensionalReconciliationCVQKD,num63_Kish2024_Mitigation_ChannelTampering_CVQKD_PRResearch}, see Figure~\ref{fig:modality-trust}.

\subsection{Deployment contexts}
\textbf{Fiber.} Experimental and field results span metropolitan and long-reach fiber links for CV-QKD, including excess-noise control and real-environment monitoring~\cite{num11_Huang2016_LongDistanceCVQKD_ExcessNoiseControl,num12_Jouguet2013_LongDistanceCVQKD_ExperimentalDemo,num72_Liu2017_Monitoring_CVQKD_RealEnvironment,num02_Hajomer2022_CVQKD60kmRLO}. 
\textbf{Free-space.} Free-space trials implement turbulence mitigation (beam steering, quadrant sensing, PID control) with reinforcement learning aiding controller tuning to stabilize the quantum link~\cite{fazzari2024controlling,guatto2024improving}.  Satellite communication provides another free-space deployment context where quantum cryptography can be combined with ML-based security monitoring~\cite{nadeem2026enhancing}. 
\textbf{Coexistence with classical optics (SWDM/SDM).} In beyond-5G fronthaul and multi-band/multi-core settings, quantum channels co-propagate with classical traffic and face Raman scattering, four-wave mixing, and inter-core crosstalk; ML (e.g., LightGBM/XGBoost) has been proposed to predict noise and reduce evaluation time for channel allocation~\cite{num78_Wang2022_NoisePrediction_ML_Secured_SWDM_B5G_Fronthaul,num79_Markowski2016_FWM_DWDM_1310nm_ApplOpt,num80_Kong2022_CoreWavelengthAllocation_SNS_QKD_MCF_OFC, num81_Niu2019_LightGBM_NoiseSuppressing_DWDM_QKD,num83_Winzer2018_FiberOptic_Transmission_Networking_20Years, num84_Napoli2018_TowardsMultibandOpticalSystems_PND,num85_Diamanti2015_DistributingSecretKeys_CV_QKD_Entropy}. \textbf{Integrated/networked settings.}
QKD is positioned within IDS-aware, secure-routing networks (WSN/MANET; IoT/industrial/rail), augmenting point-to-point confidentiality under practical resource and latency constraints~\cite{num99_Alahmadi2022_CyberSecurity_Threats_SideChannel_DigitalAgriculture_Sensors,num100_Chkirbene2020_TIDCS_DynamicIDS_FeatureSelection_Access,num101_Hijazi2018_DL_IDS_IndustryNetwork_BDCSIntell,num102_Vacca2012_ComputerInformationSecurityHandbook,num103_Liao2013_IDS_ComprehensiveReview_JNCA,num104_Tama2019_TSEIDS_TwoStageClassifier_AnomalyIDS_Access,num105_GarciaTeodoro2009_AnomalyBased_IDS_Survey_CompSec,num106_Gumusbas2021_Survey_Databases_DL_Cybersecurity_IDS_IEEE_SystemsJ,num107_Mighan2018_DL_LatentFeatureExtraction_IDS_ICEE,num108_Ma2022_ESCVAD_EnergySavingRouting_Voronoi_WSN_IoTJ,num109_Sirisha2023_EfficientMLTechniques_IDS_ICSCDS,num110_Srilakshmi2022_SecureOptimizationRouting_MANETs_Access,num111_Hung2020_EnergyEfficient_CooperativeRouting_Heterogeneous_WSN_Access,num46_AlMohammed2021_GLOBECOM_NN_QKD_IoT, num69_AlMohammed2021_OnUseQuantumComm_SecuringIoT_6G_ICCWS,num70_AlMohammed2021_ML_DetectAttackers_QKD_IoT_Railway_IEEEAccess}.  QKD applications to IoT security motivate the inclusion of resource-constrained IoT nodes, gateways, and industrial controllers in the deployment taxonomy~\cite{al2021quantum}

\subsection{Threat model}
We adopt a problem-driven threat model organized not by adversary capability (e.g., individual vs.\ collective attacks) but by \emph{where} the adversary acts, because the locus of action determines which countermeasure family can plausibly respond and how its effect must be quantified. The four loci we use are device side-channels, channel/coexistence manipulation, protocol/estimation processes, and the ML pipeline itself when ML is in the loop. The defender's evaluation question at each locus is the same: \emph{how much detection probability is bought for how much SKR, distance, and latency?}

\begin{itemize}
  \item \textbf{Device side-channels (DV and CV).} 
  For CV-QKD: local-oscillator fluctuation, wavelength, calibration, saturation, and homodyne-detector blinding attacks~\cite{num05_Ma2013_LOFluctuationLoophole_CVQKD,num06_Huang2013_WavelengthAttack_CVQKD,num07_Ma2013_WavelengthAttack_Heterodyne_CVQKD,num08_Jouguet2013_PreventingCalibrationAttacks_LO_CVQKD,num09_Qin2016_SaturationAttack_CVQKD,num10_Qin2018_HomodyneBlinding_CVQKD}. 
  For DV-QKD: bright-illumination, after-gate, time-shift, and phase-remapping attacks~\cite{num14_Lydersen2010_HackingCommercialQKD_BrightIllumination,num48_Wiechers2011_AfterGateAttack_NJP,num50_Qi2007_TimeShiftAttack_QIC,num97_Fung2007_PhaseRemappingAttack_PRA}.
  
  \item \textbf{Channel manipulation/coexistence.} 
  Physical-layer tampering, including channel amplification that biases parameter estimation and reduces SKR~\cite{num63_Kish2024_Mitigation_ChannelTampering_CVQKD_PRResearch}; and impairment from co-propagating classical signals in SWDM/SDM systems (Raman, four-wave mixing, inter-core crosstalk)~\cite{num78_Wang2022_NoisePrediction_ML_Secured_SWDM_B5G_Fronthaul,num79_Markowski2016_FWM_DWDM_1310nm_ApplOpt,num80_Kong2022_CoreWavelengthAllocation_SNS_QKD_MCF_OFC,num81_Niu2019_LightGBM_NoiseSuppressing_DWDM_QKD,zitelli2025thermodynamic}.
  
  \item \textbf{Protocol/estimation issues.} 
  Sifting and high-throughput processing can leave abnormal fragments whose patterns are exploitable; ML-assisted sifting with short samples has been proposed to flag disturbed signals while consuming only a marginal fraction of raw keys~\cite{num96_Scarani2009_Security_PracticalQKD_RMP,num97_Fung2007_PhaseRemappingAttack_PRA,num98_Wu2022_SiftingScheme_CVQKD_ShortSamples_JOSAB,martinez2025sample}.
  
  \item \textbf{ML-specific risks (when ML is used).} 
  Adversarial examples can degrade CV-QKD attack detectors, motivating adversarial-perturbation elimination (e.g., GAN-based defenses)~\cite{num31_Du2022_MultiAttackDetection_NN_CVQKD,num32_Tang2023_APE_GAN_Defense_CVQKD}; training-data integrity and robustness under distribution shift are critical when ML is in the loop.
\end{itemize}

When ML/QML blocks are introduced, additional risks arise (kernel poisoning, PQC/VQA backdoors, adversarial qubit noise)~\cite{num115_Nguyen_6G_QML_AdversarialThreats_IEEE}, creating new attack surfaces that must be considered in system design and evaluation.

\section{Taxonomy of Practical Problems}
\label{sec:taxonomy}

\begin{figure}[!t]
\centering
\includegraphics[width=3.5in]{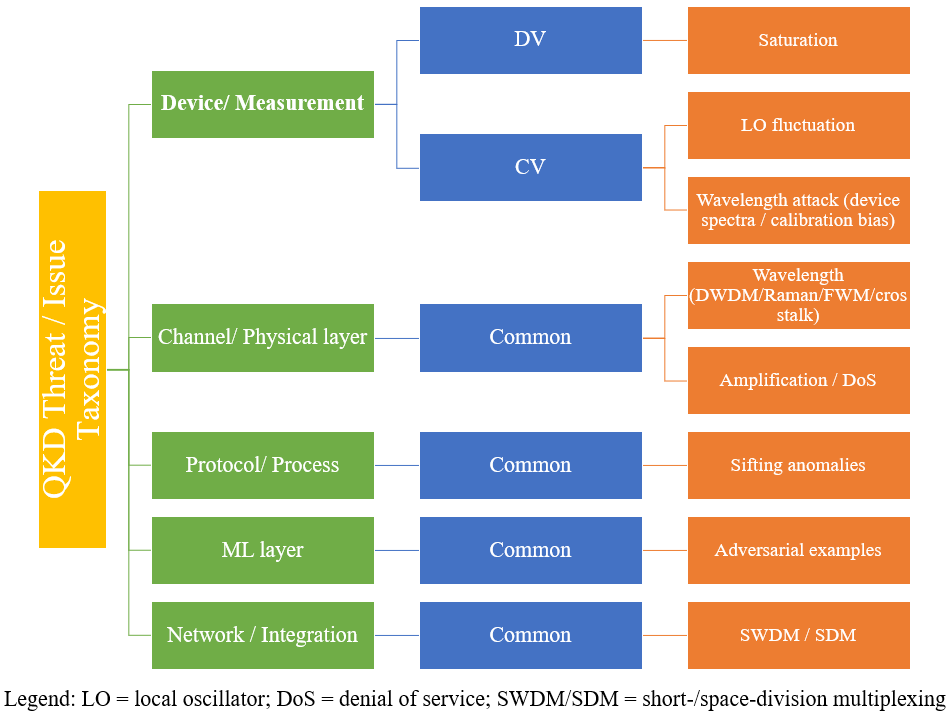}
  \caption{QKD threat/issue taxonomy.}
  \label{fig:taxonomy}
\end{figure}

This survey organizes practical QKD issues across five horizontally arranged layers—\emph{device/measurement}, \emph{channel/physical}, \emph{protocol/process}, \emph{ML}, and \emph{network/integration}—chosen so that each problem class is owned by exactly one layer for the purpose of analysis, while acknowledging that real attacks frequently cross layers (e.g., a wavelength manipulation attack is technically a device-layer side-channel but its detection signature is observed in protocol-layer estimation statistics). For each problem class we later apply the same template—\emph{Problem} $\rightarrow$ \emph{Classical solution} $\rightarrow$ \emph{ML solution} $\rightarrow$ \emph{Comparative analysis} $\rightarrow$ \emph{Gaps}—covering DV and CV in parallel; the hierarchical view is shown in Figure~\ref{fig:taxonomy}.

\subsection{Device/Measurement Side-Channels (DV and CV)}
Imperfections in sources, modulators, and detectors introduce exploitable leakages.
\begin{itemize}
    \item \textbf{CV-specific:} local-oscillator fluctuation, wavelength, calibration, saturation, and homodyne-detector blinding attacks~\cite{num05_Ma2013_LOFluctuationLoophole_CVQKD,num06_Huang2013_WavelengthAttack_CVQKD,num07_Ma2013_WavelengthAttack_Heterodyne_CVQKD, num08_Jouguet2013_PreventingCalibrationAttacks_LO_CVQKD,num09_Qin2016_SaturationAttack_CVQKD,num10_Qin2018_HomodyneBlinding_CVQKD}.
    \item \textbf{DV-specific:} time-shift, after-gate, and related bright-illumination/Trojan-horse style vectors \cite{num96_Scarani2009_Security_PracticalQKD_RMP,num14_Lydersen2010_HackingCommercialQKD_BrightIllumination,num48_Wiechers2011_AfterGateAttack_NJP,num50_Qi2007_TimeShiftAttack_QIC}.
\end{itemize}
\textit{Survey pointers.} Background and system-level perspectives appear in \cite{num04_Gisin2002_QuantumCryptographyReview, num29_Li2017_CVQKD_ChinesePhysicsB_Review, num96_Scarani2009_Security_PracticalQKD_RMP, num13_Xu2020_SecureQKDRealisticDevices}.

\subsection{Channel / Physical-Layer Manipulation}
Adversaries can bias estimation or disrupt operation at the optical-layer.
\begin{itemize}
    \item \textbf{Amplification / DoS / tampering:} channel-level manipulations that skew parameters and impair SKR~\cite{num63_Kish2024_Mitigation_ChannelTampering_CVQKD_PRResearch}.
    \item \textbf{Free-space:} turbulence and pointing/tracking challenges (link stability and estimation robustness) \cite{num13_Xu2020_SecureQKDRealisticDevices}, with modality-agnostic implications for DV and CV.
\end{itemize}

\subsection{Protocol-Level Issues}
ML-based analysis of QKD key-length choices is relevant to the protocol/process layer because key length, security margin, and post-processing cost jointly affect practical SKR and robustness~\cite{al2024impact}. 
\begin{itemize}
    \item \textbf{Parameter estimation drift \& finite-size effects:} composable finite-key and reduction results (CV)~\cite{num74_Furrer2012_FiniteKey_ComposableSecurity_CoherentAttacks_PRL,num75_Leverrier2017_Security_CVQKD_GaussianDeFinetti}; practical estimation under device imperfections and bandwidth limits~\cite{num96_Scarani2009_Security_PracticalQKD_RMP,num13_Xu2020_SecureQKDRealisticDevices,num47_Jouguet2012_Imperfections_Practical_CVQKD,num71_Wang2016_FiniteSamplingBandwidth_CVQKD}.
    \item \textbf{Sifting anomalies:} short-sample sifting and screening for disturbed signals~\cite{num98_Wu2022_SiftingScheme_CVQKD_ShortSamples_JOSAB}; interactions with throughput and data retention.
    \item \textbf{Reconciliation/latency trade-offs:} reconciliation efficiency $\beta$ and coding choices impact SKR and delay; multidimensional reconciliation and Bayesian improvements illustrate operating-point sensitivities~\cite{num03_Leverrier2008_MultidimensionalReconciliationCVQKD,num86_Kleis2019_Improving_SKR_BayesianInference_CoherentQKD}. At the post-processing layer, ML-assisted Cascade reconciliation has been explored as a route toward scalable QKD, linking reconciliation efficiency directly to SKR, latency, and practical deployment feasibility~\cite{al2024towards}.
    \item \textbf{Authentication cost:} authenticated classical exchanges are assumed and consume budget in practical deployments \cite{num13_Xu2020_SecureQKDRealisticDevices}.
\end{itemize}

\subsection{ML-Layer Threats (When ML is in the Loop)}
ML aids configuration, monitoring, and defense, but introduces its own risks.
\begin{itemize}
    \item \textbf{Detector brittleness \& generalization:} supervised detectors for attack/quality monitoring face distribution shift and class imbalance~\cite{num30_Mao2020_ML_DetectingQuantumAttacks_CVQKD,num31_Du2022_MultiAttackDetection_NN_CVQKD, num37_Liao2022_DBSCAN_DetectingAttacks_CVQKD}.
    \item \textbf{Adversarial examples \& defenses:} one-pixel attacks and learned perturbations degrade CV-QKD detectors; defenses include GAN-based purification and robust manifold methods~\cite{num32_Tang2023_APE_GAN_Defense_CVQKD, num58_Fu2023_KRMNMF_Defense_CVQKD,num27_Guo2023_OnePixelAttack_CVQKD}. Broader adversarial-learning context appears in~\cite{num61_Lu2020_QuantumAdversarialML_PRResearch,num59_Ren2022_ExperimentalQuantumAdversarialLearning}, and CV-tailored artificial key fingerprints in~\cite{ num60_Yan2023_ArtificialKeyFingerprints_CVQKD}.
\end{itemize}

\subsection{Network / Integration}
End-to-end security depends on coexistence, operations, and resource control beyond a single link.
\begin{itemize}
    \item \textbf{Coexistence with classical traffic (SWDM/SDM/multiband):} Raman, four-wave mixing, and inter-core crosstalk motivate ML-based noise prediction and channel/core allocation~\cite{num78_Wang2022_NoisePrediction_ML_Secured_SWDM_B5G_Fronthaul,num79_Markowski2016_FWM_DWDM_1310nm_ApplOpt,num80_Kong2022_CoreWavelengthAllocation_SNS_QKD_MCF_OFC, num81_Niu2019_LightGBM_NoiseSuppressing_DWDM_QKD,num83_Winzer2018_FiberOptic_Transmission_Networking_20Years, num84_Napoli2018_TowardsMultibandOpticalSystems_PND,num85_Diamanti2015_DistributingSecretKeys_CV_QKD_Entropy}.
    \item \textbf{IDS integration:} classical intrusion-detection (signature/anomaly/ensembles) informs monitoring around QKD components and controllers~\cite{num99_Alahmadi2022_CyberSecurity_Threats_SideChannel_DigitalAgriculture_Sensors,num100_Chkirbene2020_TIDCS_DynamicIDS_FeatureSelection_Access,num101_Hijazi2018_DL_IDS_IndustryNetwork_BDCSIntell,num102_Vacca2012_ComputerInformationSecurityHandbook,num103_Liao2013_IDS_ComprehensiveReview_JNCA,num104_Tama2019_TSEIDS_TwoStageClassifier_AnomalyIDS_Access,num105_GarciaTeodoro2009_AnomalyBased_IDS_Survey_CompSec,num106_Gumusbas2021_Survey_Databases_DL_Cybersecurity_IDS_IEEE_SystemsJ,num109_Sirisha2023_EfficientMLTechniques_IDS_ICSCDS,num110_Srilakshmi2022_SecureOptimizationRouting_MANETs_Access,num111_Hung2020_EnergyEfficient_CooperativeRouting_Heterogeneous_WSN_Access}.
    \item \textbf{QKD-network resource allocation:} multi-tenant scheduling and trusted/untrusted relay design (with heuristics/ML) for backbone and metro contexts~\cite{num01_Cao2020_MultiTenantProvisioningQKDNet,num95_Cao2021_HybridTrustedUntrustedRelay_QKD_Backbone_JSAC}; privacy-preserving resource learning at the edge~\cite{num94_Xu2023_PrivacyPreserving_ResourceAllocation_FederatedEdge_QI_JSTSP}.
\end{itemize}

Table~\ref{tab:threat-map} summarizes representative threats and mitigations across the five attack surfaces, contrasting DV and CV approaches and listing their dominant SKR/latency cost. Reading the table column-wise reveals a structural asymmetry between DV and CV: DV-side vulnerabilities concentrate sharply on the detector (after-gate, time-shift, bright-illumination)~\cite{num48_Wiechers2011_AfterGateAttack_NJP,num50_Qi2007_TimeShiftAttack_QIC}, so MDI/DI shifts that move trust off the measurement device are an especially natural response. CV-side vulnerabilities are more distributed across the LO branch, calibration, and saturation regimes~\cite{num05_Ma2013_LOFluctuationLoophole_CVQKD,num06_Huang2013_WavelengthAttack_CVQKD,num07_Ma2013_WavelengthAttack_Heterodyne_CVQKD,num08_Jouguet2013_PreventingCalibrationAttacks_LO_CVQKD,num09_Qin2016_SaturationAttack_CVQKD,num10_Qin2018_HomodyneBlinding_CVQKD}, which makes a single architectural fix less likely to be a full solution and which in turn explains why most ML detection work has concentrated on the CV side—the residual attack surface after physical hardening is broader and more multidimensional, exactly the regime where data-driven classifiers add value. Channel impairments such as four-wave mixing and adversarial tampering benefit from ML-based noise prediction and mitigation~\cite{num79_Markowski2016_FWM_DWDM_1310nm_ApplOpt,num63_Kish2024_Mitigation_ChannelTampering_CVQKD_PRResearch,num78_Wang2022_NoisePrediction_ML_Secured_SWDM_B5G_Fronthaul}; at the protocol level, finite-key security still rests primarily on composable bounds and Gaussian de Finetti reductions~\cite{num74_Furrer2012_FiniteKey_ComposableSecurity_CoherentAttacks_PRL,num75_Leverrier2017_Security_CVQKD_GaussianDeFinetti}, with ML acting as a real-time companion rather than a replacement; ML-layer defenses (APE-GAN, robust preprocessing) counter adversarial examples~\cite{num32_Tang2023_APE_GAN_Defense_CVQKD,num27_Guo2023_OnePixelAttack_CVQKD}; and network coexistence planning uses ML-assisted resource allocation and noise suppression for DWDM, MCF, and SWDM scenarios~\cite{num80_Kong2022_CoreWavelengthAllocation_SNS_QKD_MCF_OFC,num81_Niu2019_LightGBM_NoiseSuppressing_DWDM_QKD,num78_Wang2022_NoisePrediction_ML_Secured_SWDM_B5G_Fronthaul}. The right-most ``SKR/Latency'' column makes the deployment-cost dimension explicit and is the column that defenses ultimately compete in.

\begin{table}[!t]
  \centering
  \scriptsize
  \setlength{\tabcolsep}{1.8pt}%
  \renewcommand{\arraystretch}{1.03}%
  \caption{Threat–mitigation map contrasting DV and CV across five attack surfaces with representative SKR/latency trade-offs.}
  \label{tab:threat-map}
  {\setlength{\emergencystretch}{2em}%
  \begin{tabular}{p{0.18\columnwidth} p{0.31\columnwidth} p{0.31\columnwidth} p{0.16\columnwidth}}
    \hline
    \textbf{\mbox{Threat row}} & \textbf{DV (exemplars)} & \textbf{CV (exemplars)} & \textbf{\mbox{SKR/Latency}} \\
    \hline
    Device &
    after-gate; time-shift~\cite{num48_Wiechers2011_AfterGateAttack_NJP,num50_Qi2007_TimeShiftAttack_QIC} &
    wavelength/LO manipulation; calibration; saturation/blinding~\cite{num05_Ma2013_LOFluctuationLoophole_CVQKD,num06_Huang2013_WavelengthAttack_CVQKD,num07_Ma2013_WavelengthAttack_Heterodyne_CVQKD,num08_Jouguet2013_PreventingCalibrationAttacks_LO_CVQKD,num09_Qin2016_SaturationAttack_CVQKD,num10_Qin2018_HomodyneBlinding_CVQKD} &
    real-time path; minimal added loss \\[2pt]
    Channel &
    Raman/crosstalk management; isolation; guard bands &
    FWM; channel tampering; ML noise prediction~\cite{num79_Markowski2016_FWM_DWDM_1310nm_ApplOpt,num63_Kish2024_Mitigation_ChannelTampering_CVQKD_PRResearch,num78_Wang2022_NoisePrediction_ML_Secured_SWDM_B5G_Fronthaul} &
    SKR hinges on SNR; sub-ms control \\[2pt]
    Protocol/Finite-key &
    tighter PE; authentication hardening; MDI/DI shift &
    composable finite-key bounds; Gaussian de Finetti~\cite{num74_Furrer2012_FiniteKey_ComposableSecurity_CoherentAttacks_PRL,num75_Leverrier2017_Security_CVQKD_GaussianDeFinetti} &
    finite-key penalty vs.\ rate \\[2pt]
    ML layer &
    robust preprocessing; sanity checks &
    adversarial examples; APE-GAN defense; retraining~\cite{num32_Tang2023_APE_GAN_Defense_CVQKD,num27_Guo2023_OnePixelAttack_CVQKD} &
    inference budget; batching \\[2pt]
    Network/Coexistence &
    DWDM layout; guard-band planning &
    MCF/SWDM planning; ML-assisted allocation/noise suppression~\cite{num80_Kong2022_CoreWavelengthAllocation_SNS_QKD_MCF_OFC,num81_Niu2019_LightGBM_NoiseSuppressing_DWDM_QKD,num78_Wang2022_NoisePrediction_ML_Secured_SWDM_B5G_Fronthaul} &
    SKR vs.\ insertion loss/latency \\
    \hline
  \end{tabular}%
  }
\end{table}

\section{Survey Methodology}
\label{sec:methodology}
This section documents how the corpus underlying the survey was assembled, tagged, normalized, and graded for reproducibility. The procedure is explicit because problem-driven surveys are only useful insofar as the comparison frame is honest about what was and was not included, and what was and was not directly comparable.

\subsection{Literature selection and inclusion criteria}
We include \emph{peer-reviewed papers and flagship conference contributions} that address QKD security (DV and CV), optical co-existence, and ML-driven monitoring/defense, complemented by clearly marked preprints when present in the corpus. Representative journals and venues in the draft include
\emph{Rev.\ Mod.\ Phys.} \cite{num96_Scarani2009_Security_PracticalQKD_RMP,num13_Xu2020_SecureQKDRealisticDevices}, \emph{Phys.\ Rev.\ Lett.}~\cite{num16_Acin2007_DeviceIndependentSecurity_CollectiveAttacks,num17_Lo2012_MDIQKD_PRL,num42_Grosshans2002_CVQKD_CoherentStates_PRL,num75_Leverrier2017_Security_CVQKD_GaussianDeFinetti}, \emph{Phys.\ Rev.\ A}~\cite{num05_Ma2013_LOFluctuationLoophole_CVQKD,num06_Huang2013_WavelengthAttack_CVQKD,num07_Ma2013_WavelengthAttack_Heterodyne_CVQKD, num08_Jouguet2013_PreventingCalibrationAttacks_LO_CVQKD,num09_Qin2016_SaturationAttack_CVQKD,num10_Qin2018_HomodyneBlinding_CVQKD,num97_Fung2007_PhaseRemappingAttack_PRA,num21_Liu2018_ML_AutomaticParameterPrediction_CVQKD,num23_Wang2019_ML_OptimalParameterPrediction_QKD,num03_Leverrier2008_MultidimensionalReconciliationCVQKD,num71_Wang2016_FiniteSamplingBandwidth_CVQKD,num60_Yan2023_ArtificialKeyFingerprints_CVQKD,num56_Zheng2019_PracticalSecurity_ReducedOpticalAttenuation_CVQKD}, \emph{Nat.\ Photonics}  \cite{num14_Lydersen2010_HackingCommercialQKD_BrightIllumination,num12_Jouguet2013_LongDistanceCVQKD_ExperimentalDemo}, \emph{Nature Communications}~\cite{num15_Gerhardt2011_PerfectEavesdropperImplementation}, \emph{Optics Express}~\cite{num72_Liu2017_Monitoring_CVQKD_RealEnvironment,num81_Niu2019_LightGBM_NoiseSuppressing_DWDM_QKD,num83_Winzer2018_FiberOptic_Transmission_Networking_20Years}, \emph{J.\ Lightwave Technol.}~\cite{num86_Kleis2019_Improving_SKR_BayesianInference_CoherentQKD,num88_Zibar2020_InverseSystemDesign_RamanAmplifier_JLT,num89_Hager2019_MultistepNonlinearityCompensation_ML_ECOC,num90_Khan2019_OpticalCommPerspective_ML_Applications_JLT,num91_Karanov2018_EndToEnd_DL_OpticalFiberComms_JLT}, \emph{IEEE Access}~\cite{num70_AlMohammed2021_ML_DetectAttackers_QKD_IoT_Railway_IEEEAccess,num100_Chkirbene2020_TIDCS_DynamicIDS_FeatureSelection_Access,num104_Tama2019_TSEIDS_TwoStageClassifier_AnomalyIDS_Access}, \emph{PRX Quantum}~\cite{num18_Wallnoefer2020_ML_LongDistanceQuantumCommunication}, \emph{Phys.\ Rev.\ Research}~\cite{num63_Kish2024_Mitigation_ChannelTampering_CVQKD_PRResearch,num27_Guo2023_OnePixelAttack_CVQKD,num61_Lu2020_QuantumAdversarialML_PRResearch}, \emph{Entropy}/\emph{Photonics}/\emph{Electronics}~\cite{num32_Tang2023_APE_GAN_Defense_CVQKD, num58_Fu2023_KRMNMF_Defense_CVQKD,num27_Guo2023_OnePixelAttack_CVQKD,num25_Su2019_BP_NN_ParameterOptimization_Atmosphere_CVQKD,num26_Huang2021_SecureCVQKD_ML}, and conference series (OFC/ECOC/ICCT/HPCA)~\cite{num02_Hajomer2022_CVQKD60kmRLO,num89_Hager2019_MultistepNonlinearityCompensation_ML_ECOC,num80_Kong2022_CoreWavelengthAllocation_SNS_QKD_MCF_OFC,num87_Zibar2020_AdvancingClassicalQuantumComms_ML_OFC,num92_Jones2019_EndToEndLearning_GMI_GeometricConstellation_ECOC}. We retain arXiv entries explicitly listed (e.g.,~\cite{num65_Saad2019_Vision6G_ArXiv,num77_Cai2019_MulticoreFiber_QKD_AccessNetwork_arXiv}).

\noindent\textbf{Inclusion.} We include works that (i) study practical threats (side-channels, channel tampering/co-existence, protocol processes)~\cite{num05_Ma2013_LOFluctuationLoophole_CVQKD,num06_Huang2013_WavelengthAttack_CVQKD,num07_Ma2013_WavelengthAttack_Heterodyne_CVQKD, num08_Jouguet2013_PreventingCalibrationAttacks_LO_CVQKD,num09_Qin2016_SaturationAttack_CVQKD,num10_Qin2018_HomodyneBlinding_CVQKD,num14_Lydersen2010_HackingCommercialQKD_BrightIllumination,num48_Wiechers2011_AfterGateAttack_NJP,num50_Qi2007_TimeShiftAttack_QIC,num96_Scarani2009_Security_PracticalQKD_RMP,num97_Fung2007_PhaseRemappingAttack_PRA,num98_Wu2022_SiftingScheme_CVQKD_ShortSamples_JOSAB,num63_Kish2024_Mitigation_ChannelTampering_CVQKD_PRResearch}; (ii) develop classical mitigations or architectural shifts (e.g., MDI/DI, calibration/excess-noise control)~\cite{num11_Huang2016_LongDistanceCVQKD_ExcessNoiseControl,num08_Jouguet2013_PreventingCalibrationAttacks_LO_CVQKD,num16_Acin2007_DeviceIndependentSecurity_CollectiveAttacks,num17_Lo2012_MDIQKD_PRL,num28_Tang2014_MDIQKD_Polarization_Experimental}; (iii) propose ML-enabled parameter prediction, monitoring, attack detection, adversarial defense, or network-level decision support~\cite{num72_Liu2017_Monitoring_CVQKD_RealEnvironment,num21_Liu2018_ML_AutomaticParameterPrediction_CVQKD,num23_Wang2019_ML_OptimalParameterPrediction_QKD,num24_Ding2020_RandomForest_ParamPrediction_QKD, num25_Su2019_BP_NN_ParameterOptimization_Atmosphere_CVQKD,num30_Mao2020_ML_DetectingQuantumAttacks_CVQKD,num31_Du2022_MultiAttackDetection_NN_CVQKD, num37_Liao2022_DBSCAN_DetectingAttacks_CVQKD,num32_Tang2023_APE_GAN_Defense_CVQKD, num58_Fu2023_KRMNMF_Defense_CVQKD,num94_Xu2023_PrivacyPreserving_ResourceAllocation_FederatedEdge_QI_JSTSP}; or (iv) characterize optical co-existence and resource allocation relevant to QKD operation~\cite{num78_Wang2022_NoisePrediction_ML_Secured_SWDM_B5G_Fronthaul,num79_Markowski2016_FWM_DWDM_1310nm_ApplOpt,num80_Kong2022_CoreWavelengthAllocation_SNS_QKD_MCF_OFC, num81_Niu2019_LightGBM_NoiseSuppressing_DWDM_QKD,num83_Winzer2018_FiberOptic_Transmission_Networking_20Years, num84_Napoli2018_TowardsMultibandOpticalSystems_PND,num85_Diamanti2015_DistributingSecretKeys_CV_QKD_Entropy,num95_Cao2021_HybridTrustedUntrustedRelay_QKD_Backbone_JSAC}.

\subsection{Tagging schema (problem class and deployment)}
Each paper is tagged along two axes to enable consistent comparisons:

\noindent\textbf{(A) Problem class (cf.\ Sec.~\ref{sec:taxonomy}).}
\begin{itemize}
    \item \emph{Device/measurement side-channels} (CV: LO fluctuation, wavelength, calibration, saturation, HD blinding; DV: time-shift/after-gate/bright-illumination/phase-remap)~\cite{num05_Ma2013_LOFluctuationLoophole_CVQKD,num06_Huang2013_WavelengthAttack_CVQKD,num07_Ma2013_WavelengthAttack_Heterodyne_CVQKD, num08_Jouguet2013_PreventingCalibrationAttacks_LO_CVQKD,num09_Qin2016_SaturationAttack_CVQKD,num10_Qin2018_HomodyneBlinding_CVQKD,num14_Lydersen2010_HackingCommercialQKD_BrightIllumination,num48_Wiechers2011_AfterGateAttack_NJP,num50_Qi2007_TimeShiftAttack_QIC,num97_Fung2007_PhaseRemappingAttack_PRA}.
    \item \emph{Channel/physical-layer manipulation} (tampering, amplification/DoS; free-space stability)~\cite{num13_Xu2020_SecureQKDRealisticDevices,num63_Kish2024_Mitigation_ChannelTampering_CVQKD_PRResearch}.
    \item \emph{Protocol/process} (parameter estimation drift, finite-size, reconciliation; sifting/latency/authentication)~\cite{num96_Scarani2009_Security_PracticalQKD_RMP,num13_Xu2020_SecureQKDRealisticDevices,num98_Wu2022_SiftingScheme_CVQKD_ShortSamples_JOSAB,num74_Furrer2012_FiniteKey_ComposableSecurity_CoherentAttacks_PRL,num75_Leverrier2017_Security_CVQKD_GaussianDeFinetti,num03_Leverrier2008_MultidimensionalReconciliationCVQKD,num86_Kleis2019_Improving_SKR_BayesianInference_CoherentQKD,num47_Jouguet2012_Imperfections_Practical_CVQKD,num71_Wang2016_FiniteSamplingBandwidth_CVQKD}.
    \item \emph{ML-layer threats} (detector brittleness, adversarial examples/defense)~\cite{num31_Du2022_MultiAttackDetection_NN_CVQKD,num32_Tang2023_APE_GAN_Defense_CVQKD, num58_Fu2023_KRMNMF_Defense_CVQKD,num61_Lu2020_QuantumAdversarialML_PRResearch,num59_Ren2022_ExperimentalQuantumAdversarialLearning, num60_Yan2023_ArtificialKeyFingerprints_CVQKD}.
    \item \emph{Network/integration} (SWDM/SDM co-propagation; IDS and routing; QKD resource allocation)~\cite{num78_Wang2022_NoisePrediction_ML_Secured_SWDM_B5G_Fronthaul,num79_Markowski2016_FWM_DWDM_1310nm_ApplOpt,num80_Kong2022_CoreWavelengthAllocation_SNS_QKD_MCF_OFC, num81_Niu2019_LightGBM_NoiseSuppressing_DWDM_QKD,num83_Winzer2018_FiberOptic_Transmission_Networking_20Years, num84_Napoli2018_TowardsMultibandOpticalSystems_PND,num99_Alahmadi2022_CyberSecurity_Threats_SideChannel_DigitalAgriculture_Sensors,num100_Chkirbene2020_TIDCS_DynamicIDS_FeatureSelection_Access,num101_Hijazi2018_DL_IDS_IndustryNetwork_BDCSIntell,num102_Vacca2012_ComputerInformationSecurityHandbook,num103_Liao2013_IDS_ComprehensiveReview_JNCA,num104_Tama2019_TSEIDS_TwoStageClassifier_AnomalyIDS_Access,num105_GarciaTeodoro2009_AnomalyBased_IDS_Survey_CompSec,num106_Gumusbas2021_Survey_Databases_DL_Cybersecurity_IDS_IEEE_SystemsJ,num109_Sirisha2023_EfficientMLTechniques_IDS_ICSCDS,num110_Srilakshmi2022_SecureOptimizationRouting_MANETs_Access,num111_Hung2020_EnergyEfficient_CooperativeRouting_Heterogeneous_WSN_Access,num01_Cao2020_MultiTenantProvisioningQKDNet,num94_Xu2023_PrivacyPreserving_ResourceAllocation_FederatedEdge_QI_JSTSP,num95_Cao2021_HybridTrustedUntrustedRelay_QKD_Backbone_JSAC}.
\end{itemize}

\noindent\textbf{(B) Deployment context.}
\begin{itemize}
    \item \emph{Link/medium:} fiber~\cite{num11_Huang2016_LongDistanceCVQKD_ExcessNoiseControl,num12_Jouguet2013_LongDistanceCVQKD_ExperimentalDemo,num72_Liu2017_Monitoring_CVQKD_RealEnvironment}; free-space (generic stability considerations reflected in system surveys) \cite{num13_Xu2020_SecureQKDRealisticDevices}; multi-core/multi-band/co-propagation optics~\cite{num78_Wang2022_NoisePrediction_ML_Secured_SWDM_B5G_Fronthaul,num79_Markowski2016_FWM_DWDM_1310nm_ApplOpt,num80_Kong2022_CoreWavelengthAllocation_SNS_QKD_MCF_OFC, num81_Niu2019_LightGBM_NoiseSuppressing_DWDM_QKD,num83_Winzer2018_FiberOptic_Transmission_Networking_20Years, num84_Napoli2018_TowardsMultibandOpticalSystems_PND,num85_Diamanti2015_DistributingSecretKeys_CV_QKD_Entropy}.
    \item \emph{Architecture:} DV/CV (TLO/LLO)~\cite{num29_Li2017_CVQKD_ChinesePhysicsB_Review,num12_Jouguet2013_LongDistanceCVQKD_ExperimentalDemo,num02_Hajomer2022_CVQKD60kmRLO}; MDI/DI~\cite{num16_Acin2007_DeviceIndependentSecurity_CollectiveAttacks,num17_Lo2012_MDIQKD_PRL, num28_Tang2014_MDIQKD_Polarization_Experimental}.
    \item \emph{System scale:} point-to-point vs.\ networked/backbone and multi-tenant settings~\cite{num01_Cao2020_MultiTenantProvisioningQKDNet,num95_Cao2021_HybridTrustedUntrustedRelay_QKD_Backbone_JSAC}.
\end{itemize}

\subsection{Comparison fields and metrics}
For each study, we extract a consistent set of fields to enable apples-to-apples summaries later in Section~V:
\begin{itemize}
    \item \textbf{Task \& model:} problem type (attack detection, parameter prediction, stabilization, co-existence forecasting, IDS, routing); model family (e.g., SVM/trees/NN/GAN/RF)~\cite{num21_Liu2018_ML_AutomaticParameterPrediction_CVQKD,num23_Wang2019_ML_OptimalParameterPrediction_QKD,num24_Ding2020_RandomForest_ParamPrediction_QKD, num25_Su2019_BP_NN_ParameterOptimization_Atmosphere_CVQKD,num30_Mao2020_ML_DetectingQuantumAttacks_CVQKD,num31_Du2022_MultiAttackDetection_NN_CVQKD, num37_Liao2022_DBSCAN_DetectingAttacks_CVQKD,num32_Tang2023_APE_GAN_Defense_CVQKD, num58_Fu2023_KRMNMF_Defense_CVQKD}, with DV/CV modality.
    \item \textbf{Data \& setting:} dataset origin (simulation, lab, field); sample size; feature type (optical/statistical/traffic); train/validation/test protocol; whether short-sample or streaming constraints apply~\cite{num72_Liu2017_Monitoring_CVQKD_RealEnvironment,num98_Wu2022_SiftingScheme_CVQKD_ShortSamples_JOSAB}.
    \item \textbf{Detection metrics:} precision/recall/F1 and ROC-type measures where reported (attack/quality detectors)~\cite{num30_Mao2020_ML_DetectingQuantumAttacks_CVQKD,num31_Du2022_MultiAttackDetection_NN_CVQKD, num37_Liao2022_DBSCAN_DetectingAttacks_CVQKD}; confusion analyses for multi-attack settings when available.
    \item \textbf{Security impact:} SKR change ($\Delta$SKR) and/or \emph{maximum distance} at a fixed secret-key threshold (when reported)~\cite{num11_Huang2016_LongDistanceCVQKD_ExcessNoiseControl,num12_Jouguet2013_LongDistanceCVQKD_ExperimentalDemo,num86_Kleis2019_Improving_SKR_BayesianInference_CoherentQKD}; qualitative assessment when only narrative results are given.
    \item \textbf{Compute/latency:} inference time, update rate, hardware notes (e.g., controller loop rates, FPGA/CPU) where available; we report as stated without re-benchmarking.
    \item \textbf{Data needs \& labeling:} labeled/weakly labeled/unsupervised; annotation burden; class imbalance handling~\cite{num30_Mao2020_ML_DetectingQuantumAttacks_CVQKD,num31_Du2022_MultiAttackDetection_NN_CVQKD, num37_Liao2022_DBSCAN_DetectingAttacks_CVQKD}.
    \item \textbf{Robustness:} adversarial stress tests and distribution-shift probes when present (e.g., one-pixel and GAN-based defenses, robust manifold methods)~\cite{num32_Tang2023_APE_GAN_Defense_CVQKD, num58_Fu2023_KRMNMF_Defense_CVQKD,num27_Guo2023_OnePixelAttack_CVQKD}, along with broader adversarial-learning context~\cite{num61_Lu2020_QuantumAdversarialML_PRResearch,num59_Ren2022_ExperimentalQuantumAdversarialLearning, num60_Yan2023_ArtificialKeyFingerprints_CVQKD}.
    \item \textbf{Standards alignment:} when papers situate assumptions vis-à-vis common security models and operational practices (e.g., those surveyed in \cite{num13_Xu2020_SecureQKDRealisticDevices}, \cite{num96_Scarani2009_Security_PracticalQKD_RMP}), we record this to gauge consistency with typical compliance baselines (ETSI/ITU-T, when explicitly discussed).
\end{itemize}

\subsection{Normalization and evidence grading}
Because reporting styles vary, we follow these rules:
\begin{enumerate}
    \item \emph{Use as-published metrics.} We do not infer missing statistics; when only accuracy or qualitative claims are given, we mark fields as “not reported”.
    \item \emph{Prefer comparable operating points.} When multiple working points are listed, we highlight those at similar SNR/launch power/distance (for optics) or similar block sizes (for protocol evaluation) to align comparisons~\cite{num11_Huang2016_LongDistanceCVQKD_ExcessNoiseControl,num12_Jouguet2013_LongDistanceCVQKD_ExperimentalDemo,num98_Wu2022_SiftingScheme_CVQKD_ShortSamples_JOSAB,num86_Kleis2019_Improving_SKR_BayesianInference_CoherentQKD}.
\end{enumerate}

\subsection{Reproducibility policy}
We record, for every ML-enabled study, whether \emph{code} and/or \emph{data} are available (public repository, upon request, or not available), the presence of an explicit train/validation/test split, and any notes on hyperparameter search or early stopping. Works with open artifacts receive a “reproducibility” tag; studies reporting adversarial defenses also receive an explicit “robustness evidence” tag if stress tests are documented~\cite{num32_Tang2023_APE_GAN_Defense_CVQKD, num58_Fu2023_KRMNMF_Defense_CVQKD,num27_Guo2023_OnePixelAttack_CVQKD,num61_Lu2020_QuantumAdversarialML_PRResearch}. Where only preprints are listed (e.g.,~\cite{num65_Saad2019_Vision6G_ArXiv,num77_Cai2019_MulticoreFiber_QKD_AccessNetwork_arXiv}), we mirror the availability statement in that entry.

\section{Problem-Centric Survey}
\label{sec:problem-centric}

This section follows a unified template for each problem instance:
\emph{(i) Problem \& evidence} (what breaks and under which conditions);
\emph{(ii) Classical solutions} (filters, monitors, calibration, protocol/architecture shifts; SKR/range/complexity trade-offs);
\emph{(iii) ML solutions} (features, models, training/data, on-clean accuracy and robustness to shift/adversaries; runtime);
\emph{(iv) Comparative analysis} (head-to-head summary);
\emph{(v) Open gaps \& research needs} (generalization, labeling burden, certified robustness, finite-key coupling).
For a consolidated view across problems, see the master table (Table~\ref{tab:master_VA_to_VF}).

\begin{table*}[t]
\footnotesize
\setlength{\tabcolsep}{3pt}
\renewcommand{\arraystretch}{1.05}
\centering
\caption{Master Comparison of Problems P1–P9.}
\label{tab:master_VA_to_VF}
\begin{tabular*}{\textwidth}{@{\extracolsep{\fill}}%
p{0.04\textwidth}%
p{0.07\textwidth}%
p{0.335\textwidth}%
p{0.16\textwidth}%
p{0.155\textwidth}%
p{0.195\textwidth}@{}}
\toprule
ID & Approach & Representative method & Metrics (as given) & $\Delta$SKR / dist.\ (or impact) & Notes (compute / robustness) \\
\midrule
P1 & Classical &
LO power monitors; wavelength filters; calibration hardening; MDI/DI shift~\cite{num11_Huang2016_LongDistanceCVQKD_ExcessNoiseControl,num12_Jouguet2013_LongDistanceCVQKD_ExperimentalDemo,num08_Jouguet2013_PreventingCalibrationAttacks_LO_CVQKD,num16_Acin2007_DeviceIndependentSecurity_CollectiveAttacks,num17_Lo2012_MDIQKD_PRL,num28_Tang2014_MDIQKD_Polarization_Experimental} &
As stated per study &
As stated &
Added hardware; insertion loss/latency; reduces trust on detector path (MDI/DI). \\
P1 & ML &
ANN/SVM/RF attack detectors; DBSCAN; APE-GAN / KRMNMF pre-processing~\cite{num21_Liu2018_ML_AutomaticParameterPrediction_CVQKD,num23_Wang2019_ML_OptimalParameterPrediction_QKD,num24_Ding2020_RandomForest_ParamPrediction_QKD,num30_Mao2020_ML_DetectingQuantumAttacks_CVQKD,num31_Du2022_MultiAttackDetection_NN_CVQKD,num37_Liao2022_DBSCAN_DetectingAttacks_CVQKD,num32_Tang2023_APE_GAN_Defense_CVQKD,num58_Fu2023_KRMNMF_Defense_CVQKD} &
DBSCAN (DADS): P=99.7\%, R=99.8\%, F1=0.998 (CV) &
As stated &
Latency vs.\ controller rate; sensitivity to distribution shift/adversaries~\cite{num27_Guo2023_OnePixelAttack_CVQKD,num61_Lu2020_QuantumAdversarialML_PRResearch}. \\
P1-H & ML (health) &
DL-based device defect identification (prep errors, detector efficiency) under harsh conditions; 99.7\% accuracy reported~\cite{num113_Sun2025_DL_DefectID_QKD} &
Accuracy 99.7\% &
Early fault flags; SKR preserved indirectly &
Targets device/measurement health layer; generalization across hardware/envs remains open. \\
\midrule
P2 & Classical &
DV timing/detector-control mitigation: shielding/monitoring, timing randomization; MDI/DI shift~\cite{num14_Lydersen2010_HackingCommercialQKD_BrightIllumination,num16_Acin2007_DeviceIndependentSecurity_CollectiveAttacks,num17_Lo2012_MDIQKD_PRL} &
— &
Reduces attack success; trust moved out of detector &
Hardware overhead; architectural change for MDI/DI. \\
P2 & ML &
Supervised/unsupervised anomaly detectors on DV telemetry (timing histograms, SPD stats)~\cite{num30_Mao2020_ML_DetectingQuantumAttacks_CVQKD,num31_Du2022_MultiAttackDetection_NN_CVQKD,num37_Liao2022_DBSCAN_DetectingAttacks_CVQKD} &
Detection P/R/F1; alarm latency &
Shorter detection time; lower false negatives &
Requires access to DV-side telemetry; robustness to drift. \\
\midrule
P3 & Classical &
Redundant monitors; conservative estimation windows; tamper-aware thresholds~\cite{num63_Kish2024_Mitigation_ChannelTampering_CVQKD_PRResearch} &
— &
SKR stability under tamper; potential sensitivity loss &
Low compute; can be conservative. \\
P3 & ML &
Trees/RF/NN tamper detectors; ML-guided post-selection~\cite{num30_Mao2020_ML_DetectingQuantumAttacks_CVQKD,num31_Du2022_MultiAttackDetection_NN_CVQKD,num37_Liao2022_DBSCAN_DetectingAttacks_CVQKD,num24_Ding2020_RandomForest_ParamPrediction_QKD} &
Decision-tree CA detection: 100\% (low noise) / 91.26\% (high noise) &
Post-selection SKR gain: +0.024 bits/pulse ($\sim$6 kbps @ 100 MHz) for CA; +1.6 kbps for CA-DoS &
Fast inference feasible; robustness to shift required. \\
\midrule
V-C & Classical &
FSM+QD+PID stabilization for turbulence/pointing (free-space)~\cite{num13_Xu2020_SecureQKDRealisticDevices} &
Stability/BER/SNR traces &
Improved estimation stability; SKR retention &
Deterministic control; hardware-tuned. \\
V-C & ML &
RL-assisted PID tuning; data-driven stabilization policies (controller-rate constrained) &
Track error; settling time; SKR under dynamics &
Faster adaptation; needs safe exploration &
Latency bounds; telemetry quality critical. \\
\midrule
P4 & Classical &
Longer blocks; multidimensional reconciliation; Bayesian inference~\cite{num03_Leverrier2008_MultidimensionalReconciliationCVQKD,num86_Kleis2019_Improving_SKR_BayesianInference_CoherentQKD} &
Block size; FER; SKR &
Higher SKR at cost of finite-key penalties/latency &
Bandwidth and block-length constraints apply. \\
P4 & ML &
SVR/ANN/RF parameter predictors; field monitoring~\cite{num72_Liu2017_Monitoring_CVQKD_RealEnvironment,num21_Liu2018_ML_AutomaticParameterPrediction_CVQKD,num23_Wang2019_ML_OptimalParameterPrediction_QKD,num24_Ding2020_RandomForest_ParamPrediction_QKD,num25_Su2019_BP_NN_ParameterOptimization_Atmosphere_CVQKD,num41_Liu2019_PhaseModulationStabilization_ML_QKD} &
Reg.\ error; SKR at operating point &
Tracks drift; improves operating point &
Streaming-friendly; needs labels; finite-key coupling open. \\
P4-M & ML and Metaheuristic &
2E-HRIO optimizer + radial-DBSCAN failure prediction; ADS-CNN side-channel classifier~\cite{num114_Rajendran2025_CVQKD_2EHRIO_ADS_CNN} &
Reconciliation efficiency $\ge 90\%$; classifier accuracy (reported) &
Increased KGR (matched-point SKR not given) &
Optimizer/inference latency not reported; finite-key coupling TBD. \\
P4-Q & Quantum-ML (DV) &
QNN/QRL integrated with BB84/B92 (QNN-BB84, QNN-B92, QNN-QRL-V.1/V.2)~\cite{QNNQRL2025} &
Acc./Prec./Rec./F1; ROC; noise robustness &
Improved key gen.\ quality &
Composable finite-key coupling, runtime, generalization open. \\
\midrule
P5 & Classical &
Fixed short-sample windows; rule-based sifting screens~\cite{num98_Wu2022_SiftingScheme_CVQKD_ShortSamples_JOSAB} &
False alarm/skip rates &
Fast disturbance flags; small raw-key cost &
Needs calibration; privacy leakage budget. \\
P5 & ML &
Lightweight anomaly screens on sub-block summaries~\cite{num30_Mao2020_ML_DetectingQuantumAttacks_CVQKD,num31_Du2022_MultiAttackDetection_NN_CVQKD,num37_Liao2022_DBSCAN_DetectingAttacks_CVQKD,num98_Wu2022_SiftingScheme_CVQKD_ShortSamples_JOSAB} &
P/R/F1; detection latency &
Early discard of bad blocks; SKR preserved &
Tune thresholds to authentication/privacy budgets. \\
\midrule
P6 & Classical &
Pre-filters; randomized smoothing (model-agnostic) &
— &
N/A (robustness aid) &
No training data needed; may reduce clean accuracy slightly. \\
P6 & ML &
Adversarial training; APE-GAN; KRMNMF~\cite{num32_Tang2023_APE_GAN_Defense_CVQKD,num58_Fu2023_KRMNMF_Defense_CVQKD,num27_Guo2023_OnePixelAttack_CVQKD,num61_Lu2020_QuantumAdversarialML_PRResearch} &
APE-GAN robust acc.\ 74.88\%; KRMNMF 71.6\%; combo 78.8–79.5\% &
Restores detector utility under attack &
Compute overhead; certify if possible. \\
\midrule
P7 & Classical &
Guard bands; static impairment-aware planning (DWDM/MCF)~\cite{num83_Winzer2018_FiberOptic_Transmission_Networking_20Years,num79_Markowski2016_FWM_DWDM_1310nm_ApplOpt,num84_Napoli2018_TowardsMultibandOpticalSystems_PND} &
— &
Lower Raman/FWM/crosstalk; SKR stability &
Conservative capacity usage. \\
P7 & ML &
Noise prediction (e.g., XGBoost/LightGBM) and allocation policies~\cite{num78_Wang2022_NoisePrediction_ML_Secured_SWDM_B5G_Fronthaul,num81_Niu2019_LightGBM_NoiseSuppressing_DWDM_QKD} &
Eval.-time $\downarrow$ up to 98.8\%; ACPR $\approx$ 100\%; MNPR $\to$ 100\% when occupancy $>$ 80\%; $\le$25\% error vs.\ actual lowest noise &
Faster placement decisions; fewer blocking events &
Transferability across topologies open. \\
\midrule
P8 & Classical &
Signature/anomaly IDS; feature engineering; ensembles~\cite{num100_Chkirbene2020_TIDCS_DynamicIDS_FeatureSelection_Access,num104_Tama2019_TSEIDS_TwoStageClassifier_AnomalyIDS_Access} &
TPR/FPR; alert latency &
Operational visibility (IT/OT) &
Signature brittleness; false positives. \\
P8 & ML &
Deep/ensemble IDS tailored to deployment (industrial/IoT/railway)~\cite{num70_AlMohammed2021_ML_DetectAttackers_QKD_IoT_Railway_IEEEAccess,num101_Hijazi2018_DL_IDS_IndustryNetwork_BDCSIntell} &
P/R/F1; ROC; latency &
Improved detection under drift &
Privacy-preserving sharing and drift handling needed. \\
\midrule
P9 & Classical &
Heuristics; trusted/untrusted relay planning~\cite{num01_Cao2020_MultiTenantProvisioningQKDNet,num95_Cao2021_HybridTrustedUntrustedRelay_QKD_Backbone_JSAC} &
Throughput/SKR, utilization &
Reconfig.\ latency (policy-driven) &
Interpretable; may underutilize capacity. \\
P9 & ML &
Learning-assisted/federated edge policies~\cite{num94_Xu2023_PrivacyPreserving_ResourceAllocation_FederatedEdge_QI_JSTSP} &
As reported; latency &
Adapts to demand; privacy via FL &
Requires telemetry; robustness to bursts/failures. \\
\bottomrule
\end{tabular*}
\end{table*}


\subsection{Device/Measurement Side-Channels (CV focus; DV parallels)}
\subsubsection*{P1. LO fluctuation, wavelength, calibration, saturation, HD-blinding (CV)}
\textbf{Problem \& evidence.}
CV-QKD implementations are vulnerable to device-level side channels including local-oscillator (LO) fluctuation, wavelength manipulation, calibration attacks, detector saturation, and homodyne-detector blinding~\cite{num05_Ma2013_LOFluctuationLoophole_CVQKD,num06_Huang2013_WavelengthAttack_CVQKD,num07_Ma2013_WavelengthAttack_Heterodyne_CVQKD, num08_Jouguet2013_PreventingCalibrationAttacks_LO_CVQKD,num09_Qin2016_SaturationAttack_CVQKD,num10_Qin2018_HomodyneBlinding_CVQKD}. These effects bias parameter estimation and can increase excess noise or hide eavesdropping, degrading SKR and distance~\cite{num13_Xu2020_SecureQKDRealisticDevices,num11_Huang2016_LongDistanceCVQKD_ExcessNoiseControl,num12_Jouguet2013_LongDistanceCVQKD_ExperimentalDemo}. See Table~\ref{tab:master_VA_to_VF}, row \textbf{P1}.

\textbf{Classical solutions.}
Optical filtering and power monitoring on LO and auxiliary ports, hardened calibration routines, detector linearity checks, and architectural shifts toward MDI/DI to reduce trust on vulnerable modules ~\cite{num11_Huang2016_LongDistanceCVQKD_ExcessNoiseControl,num08_Jouguet2013_PreventingCalibrationAttacks_LO_CVQKD,num16_Acin2007_DeviceIndependentSecurity_CollectiveAttacks,num17_Lo2012_MDIQKD_PRL,num28_Tang2014_MDIQKD_Polarization_Experimental}. Pros: transparent physical mitigations and architectural risk reduction; cons: added insertion loss/latency, calibration overhead, and possible SKR penalties. (Table~\ref{tab:master_VA_to_VF}, row \textbf{P1}, “Classical”.)

\textbf{ML solutions.}
Supervised attack/anomaly detectors using ANN/SVM/RF with optical/statistical features; density-based clustering (DBSCAN) for unsupervised attack discovery~\cite{num21_Liu2018_ML_AutomaticParameterPrediction_CVQKD,num23_Wang2019_ML_OptimalParameterPrediction_QKD,num24_Ding2020_RandomForest_ParamPrediction_QKD,num30_Mao2020_ML_DetectingQuantumAttacks_CVQKD,num31_Du2022_MultiAttackDetection_NN_CVQKD, num37_Liao2022_DBSCAN_DetectingAttacks_CVQKD}. Robust pre-processing via adversarial perturbation elimination (GAN-based) and robust manifold reconstructions for CV-QKD detectors~\cite{num32_Tang2023_APE_GAN_Defense_CVQKD, num58_Fu2023_KRMNMF_Defense_CVQKD}. Pros: adaptive detection under time-varying conditions; cons: data/label requirements, distribution shift sensitivity, adversarial brittleness~\cite{num27_Guo2023_OnePixelAttack_CVQKD,num61_Lu2020_QuantumAdversarialML_PRResearch}.  Recent CV work embeds device-health/attack classification directly in the pipeline (e.g., ADS-CNN) alongside parameter optimizers, offering a joint path for defect and side-channel mitigation \cite{num114_Rajendran2025_CVQKD_2EHRIO_ADS_CNN}. (Table~\ref{tab:master_VA_to_VF}, rows \textbf{P1} and \textbf{P4-M}.)

\begin{figure}[!t]
  \centering
  \includegraphics[width=\columnwidth]{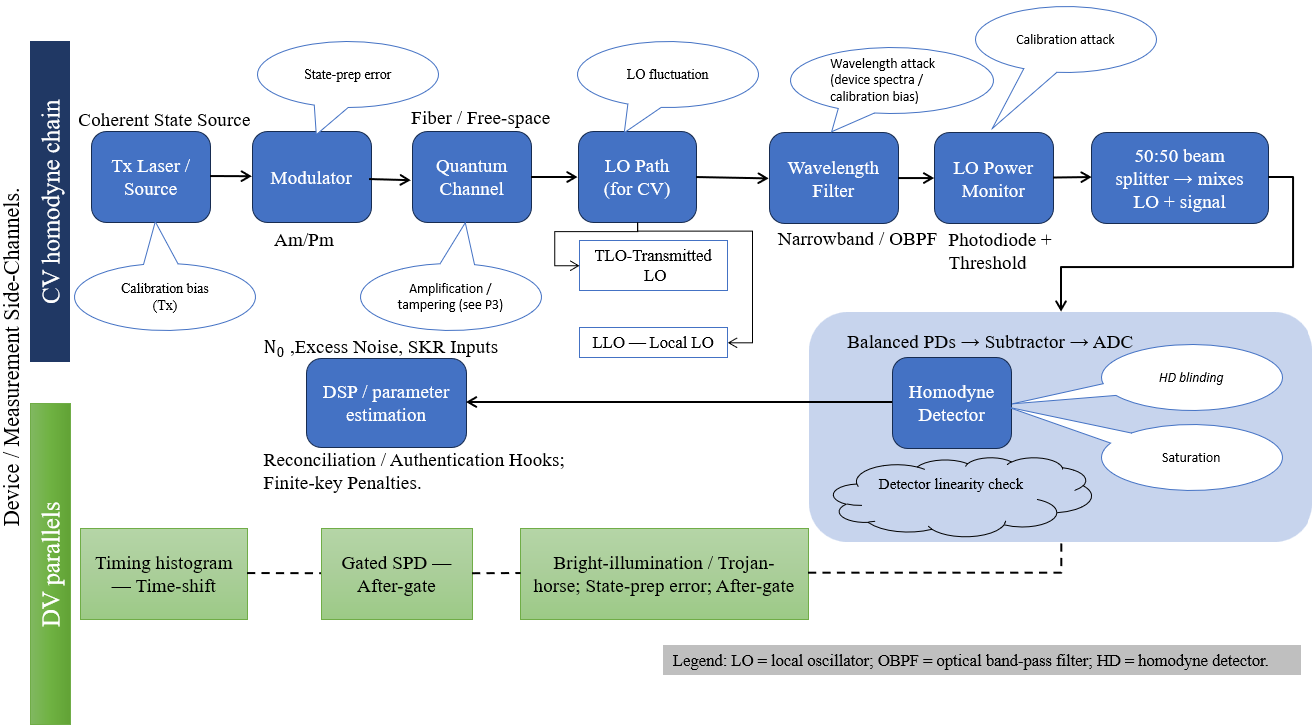}
  \caption{Device/measurement side-channels: CV homodyne chain (LO, filtering/monitoring,
  50:50 combiner, homodyne, DSP) with DV timing/illumination parallels.}
  \label{fig:sidechannels}
\end{figure}

\textbf{Deep learning for device defect identification (CV/DV-agnostic)}
Beyond attack/anomaly detectors, deep learning has been used to identify \emph{device defects}—e.g., quantum state preparation errors and degraded detector efficiency—under harsh operating conditions (EMI, temperature/humidity variations). Sun \textit{et al.} report a DL-based monitoring framework for QKD terminal equipment that attains 99.7\% identification accuracy in complex environments \cite{num113_Sun2025_DL_DefectID_QKD}. This targets the \emph{device/measurement health} layer: it complements physical monitors by flagging incipient faults that would otherwise manifest as excess noise or SKR decline. Open questions include (i) generalization across hardware and environmental domains, (ii) coupling defect alarms to finite-key post-selection thresholds, and (iii) reporting of SKR/distance and inference latency at matched operating points. As shown in Fig.~\ref{fig:sidechannels}, the CV homodyne receive chain—with LO branch (TLO/LLO), wavelength filtering and power monitoring feeding a 50:50 beam splitter, followed by the homodyne detector and DSP/parameter estimation—highlights device/measurement side-channels, while DV timing/illumination attacks (time-shift, after-gate, bright-illumination/Trojan-horse) are depicted as parallels mapped to the measurement region. (Table~\ref{tab:master_VA_to_VF}, row \textbf{P1-H}.)

\textbf{Comparative analysis.}
A side-by-side view of P1 (Table~\ref{tab:master_VA_to_VF}, rows \textbf{P1} and \textbf{P1-H}) reveals three regularities of the device-layer race in CV-QKD. \emph{First}, classical physical countermeasures and ML detectors are not substitutes but operate at different points in the kill chain: filters, power monitors, and calibration hardening reduce the \emph{probability} that an attack succeeds at altering the measured statistics, while supervised/unsupervised ML detectors reduce the \emph{time} an attack remains undetected once it has succeeded in altering them. The DBSCAN-based DADS detector at P=99.7\%/R=99.8\%/F1=0.998~\cite{num37_Liao2022_DBSCAN_DetectingAttacks_CVQKD} and the Sun \textit{et al.} DL device-health monitor at 99.7\% accuracy under EMI/environmental stress~\cite{num113_Sun2025_DL_DefectID_QKD} both illustrate that the ML side of this race is no longer the bottleneck under in-distribution conditions; the bottleneck is now distribution shift and adversarial pressure rather than nominal accuracy. \emph{Second}, MDI/DI shifts dominate the classical side asymptotically (they remove the trust assumption rather than re-establishing it through monitoring) but charge SKR and complexity, so most fielded systems will operate in the hardened-link regime where ML detectors carry real weight. \emph{Third}, joint architectures that fold defect identification and side-channel classification into the same pipeline alongside parameter optimization (e.g., ADS-CNN with 2E-HRIO~\cite{num114_Rajendran2025_CVQKD_2EHRIO_ADS_CNN}) point toward an integrated future, but the corpus is missing the matched-operating-point $\Delta$SKR and inference-latency measurements that would let one judge whether the integration is net-positive for deployment.

\textbf{Open gaps \& research needs.}
(i) Cross-setup generalization and label scarcity; (ii) explicit coupling of detector decisions to finite-key analysis; (iii) standardized stress tests for adversarial/shift robustness; (iv) quantifying SKR/distance impact with and without mitigation across identical operating points~\cite{num13_Xu2020_SecureQKDRealisticDevices,num11_Huang2016_LongDistanceCVQKD_ExcessNoiseControl,num12_Jouguet2013_LongDistanceCVQKD_ExperimentalDemo,num74_Furrer2012_FiniteKey_ComposableSecurity_CoherentAttacks_PRL,num75_Leverrier2017_Security_CVQKD_GaussianDeFinetti,num03_Leverrier2008_MultidimensionalReconciliationCVQKD}.

\subsubsection*{P2. Time-shift / after-gate / bright illumination / Trojan-horse (DV)}
\textbf{Problem \& evidence.}
DV systems face timing and detector-control attacks such as time-shift, after-gate, and bright illumination; phase-remapping has also been demonstrated \cite{num14_Lydersen2010_HackingCommercialQKD_BrightIllumination,num48_Wiechers2011_AfterGateAttack_NJP,num50_Qi2007_TimeShiftAttack_QIC,num97_Fung2007_PhaseRemappingAttack_PRA}. See Table~\ref{tab:master_VA_to_VF}, row \textbf{P2}.

\textbf{Classical solutions.}
Enhanced detector shielding and monitoring, timing randomization, and architectural shifts to MDI/DI to mitigate detector trust \cite{num14_Lydersen2010_HackingCommercialQKD_BrightIllumination,num16_Acin2007_DeviceIndependentSecurity_CollectiveAttacks,num17_Lo2012_MDIQKD_PRL}. (Table~\ref{tab:master_VA_to_VF}, row \textbf{P2}, “Classical”.)

\textbf{ML solutions.}
Where DV telemetry is available, supervised anomaly detectors mirror CV counterparts to flag distribution shifts indicative of timing or illumination manipulation (methods as in P1)~\cite{num30_Mao2020_ML_DetectingQuantumAttacks_CVQKD,num31_Du2022_MultiAttackDetection_NN_CVQKD, num37_Liao2022_DBSCAN_DetectingAttacks_CVQKD} (modality-agnostic). (Table~\ref{tab:master_VA_to_VF}, row \textbf{P2}, “ML”.)

\textbf{Comparative analysis.}
The DV picture differs from the CV one in an instructive way: classical countermeasures (shielding, timing randomization, MDI/DI) target the \emph{mechanism} of the attack at the hardware layer, while the ML literature for DV remains comparatively thin and is largely a transplant of CV-trained methodologies operating on DV telemetry (timing histograms, SPD count rates)~\cite{num30_Mao2020_ML_DetectingQuantumAttacks_CVQKD,num31_Du2022_MultiAttackDetection_NN_CVQKD, num37_Liao2022_DBSCAN_DetectingAttacks_CVQKD}. The practical consequence is that, for DV systems, ML monitors are most credible when used as drift detectors and shorteners of detection latency rather than as primary defenses—false-negative cost is high because a successful detector-blinding attack can in principle leak the entire key—so joint deployment with classical hardware mitigations and conservative post-selection remains the responsible path. The corpus does not yet contain a controlled DV adversarial-stress study analogous to the CV one-pixel/APE-GAN line, and that absence is itself the most informative comparative observation.

\textbf{Open gaps.}
Unified DV datasets for ML benchmarking; evaluation under controlled bright-illumination/time-shift campaigns; integration with finite-key and authentication budgets \cite{num96_Scarani2009_Security_PracticalQKD_RMP}.

\subsection{Channel / Physical-Layer Manipulation}
\subsubsection*{P3. Channel amplification \& (hybrid) DoS}
\begin{figure}[!t]
  \centering
  \includegraphics[width=\columnwidth]{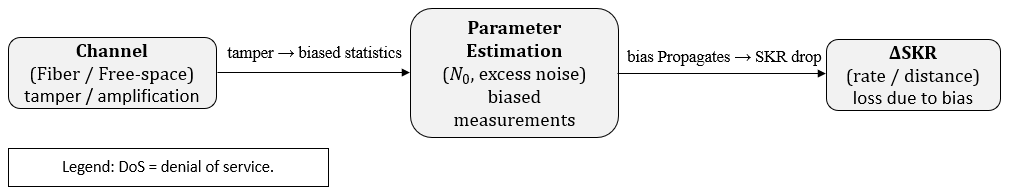}
\caption{Channel tampering/DoS: bias pathway from optical manipulation to parameter estimation and SKR impact.}
\label{fig:channel-tamper}
\end{figure}

\textbf{Problem \& evidence.}
Adversarial channel tampering or amplification can bias parameter estimation and reduce SKR, constituting a practical attack surface~\cite{num63_Kish2024_Mitigation_ChannelTampering_CVQKD_PRResearch}. See Table~\ref{tab:master_VA_to_VF}, row \textbf{P3}, and Fig.~\ref{fig:channel-tamper}

\textbf{Classical solutions.}
Redundancy in monitoring channels, conservative estimation windows, and tamper-aware thresholds; architectural diversity across links to avoid common-mode failure. (Table~\ref{tab:master_VA_to_VF}, row \textbf{P3}, “Classical”.)

\textbf{ML solutions.}
Lightweight classifiers (trees/ensembles) to detect tampering signatures; post-selection strategies guided by ML detectors to recover SKR on disturbed blocks~\cite{num30_Mao2020_ML_DetectingQuantumAttacks_CVQKD,num31_Du2022_MultiAttackDetection_NN_CVQKD, num37_Liao2022_DBSCAN_DetectingAttacks_CVQKD,num24_Ding2020_RandomForest_ParamPrediction_QKD}. (Table~\ref{tab:master_VA_to_VF}, row \textbf{P3}, “ML”.)

\textbf{Comparative analysis.}
The P3 row of Table~\ref{tab:master_VA_to_VF} contains some of the most directly interpretable numbers in the corpus: decision-tree classifiers reach 100\% channel-amplification detection in the low-noise regime but degrade to 91.26\% at high noise, and the post-selection strategy guided by such a detector contributes $\sim$+0.024 bits/pulse ($\approx 6$ kbps at 100~MHz) for the channel-amplification case and $\sim$+1.6 kbps for the joint CA-DoS case~\cite{num63_Kish2024_Mitigation_ChannelTampering_CVQKD_PRResearch,num30_Mao2020_ML_DetectingQuantumAttacks_CVQKD,num31_Du2022_MultiAttackDetection_NN_CVQKD,num37_Liao2022_DBSCAN_DetectingAttacks_CVQKD,num24_Ding2020_RandomForest_ParamPrediction_QKD}. Two observations follow. \emph{First}, the noise-regime gap (100\% $\rightarrow$ 91.26\%) is not a small effect: it is roughly an order-of-magnitude increase in false-negative rate, and it is precisely the regime in which a tampering adversary would prefer to operate, which means evaluating only at low noise will systematically overstate field performance. \emph{Second}, the SKR-gain numbers are small in absolute terms but \emph{positive}: classical conservative-window approaches typically have to discard disturbed blocks outright, so ML-guided post-selection's contribution is best read as net-recovered SKR that classical schemes leave on the floor. The cost is the inference-latency budget that the controller must absorb in line, which the corpus does not yet report at a common operating point.

\textbf{Open gaps.}
Ground-truth datasets for tampering scenarios; alignment of detector thresholds with composable security; runtime budgeting for in-line screening.

\subsection{ Free-Space Turbulence \& Pointing}
\textbf{Problem \& evidence.}
Free-space links face turbulence, beam wander, and pointing/tracking challenges that affect estimation stability and SKR~\cite{al2024tradeoffs}. High-speed railway links represent an important free-space deployment case~\cite{ALMOHAMMED2023100634}, where FSO-assisted QKD has been proposed to secure train-to-ground communication under mobility, alignment, and availability constraints \cite{num13_Xu2020_SecureQKDRealisticDevices, al2026advancing}. See Table~\ref{tab:master_VA_to_VF}, rows \textbf{V-C}.

\textbf{Classical solutions.}
Fast-steering mirror (FSM) with quadrant detector (QD) and PID control loops (generic stabilization pattern). (Table~\ref{tab:master_VA_to_VF}, \textbf{V-C}, “Classical”.)

\begin{figure}[!t]
  \centering
  \includegraphics[width=\columnwidth]{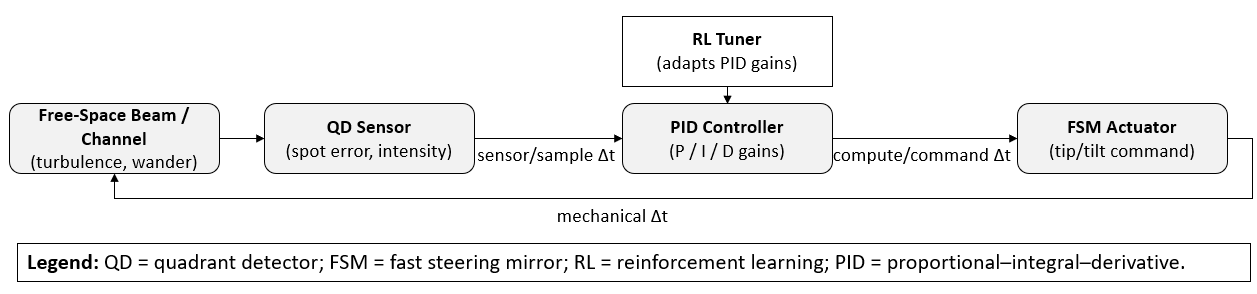}
  \caption{Free-space stabilization: FSM–QD–PID closed loop with RL-assisted PID tuning; latency and sampling points indicated.}
  \label{fig:freespace}
\end{figure}

\textbf{ML solutions.}
RL-assisted PID tuning and data-driven stabilization policies; model choice depends on controller-rate constraints and available telemetry. (Table~\ref{tab:master_VA_to_VF}, \textbf{V-C}, “ML”.), See Table~\ref{tab:master_VA_to_VF}, rows V-C (Classical/ML), See Fig.~\ref{fig:freespace} for the free-space control loop and where RL tunes the PID.

\textbf{Comparative analysis.}
The free-space turbulence/pointing problem is the cleanest test case in this survey for the \emph{classical-vs.-ML control} question, because the underlying control objective—keep the beam on the receiver aperture—admits both a textbook deterministic solution (FSM/QD/PID) and a learned one (RL-tuned PID, data-driven policies). Classical loops deliver provable stability margins, deterministic latency, and forensic transparency, which are powerful properties when something fails. RL-tuned controllers, by contrast, can adapt their gains to nonstationary turbulence statistics and recover faster after large excursions, but their certification story is weaker and their failure modes are less interpretable. The right framing is therefore not \emph{either/or} but \emph{controller + supervisor}: classical PID handles the inner loop with bounded latency, while an ML supervisor adjusts gains on a slower timescale and reverts to a known-safe schedule on uncertainty or excessive prediction error. The corpus lacks SKR-under-dynamics measurements at matched turbulence statistics for these two regimes, which is the missing measurement that would convert the design intuition above into a quantitative recommendation.

\textbf{Open gaps.}
Shared telemetry traces for benchmarking; stability proofs for ML-in-the-loop control; SKR/latency trade-off characterization.

\subsection{ Protocol-Level Issues}
\subsubsection*{P4. Parameter estimation drift \& finite-size}
\textbf{Problem \& evidence.}
Finite-size/composable analyses bound secure rates under realistic sampling; practical systems also experience bandwidth-limited estimation and imperfections~\cite{num96_Scarani2009_Security_PracticalQKD_RMP,num13_Xu2020_SecureQKDRealisticDevices,num74_Furrer2012_FiniteKey_ComposableSecurity_CoherentAttacks_PRL,num75_Leverrier2017_Security_CVQKD_GaussianDeFinetti,num47_Jouguet2012_Imperfections_Practical_CVQKD,num71_Wang2016_FiniteSamplingBandwidth_CVQKD}. See Table~\ref{tab:master_VA_to_VF}, row \textbf{P4}.

\begin{figure}[!t]
  \centering
  \includegraphics[width=\columnwidth]{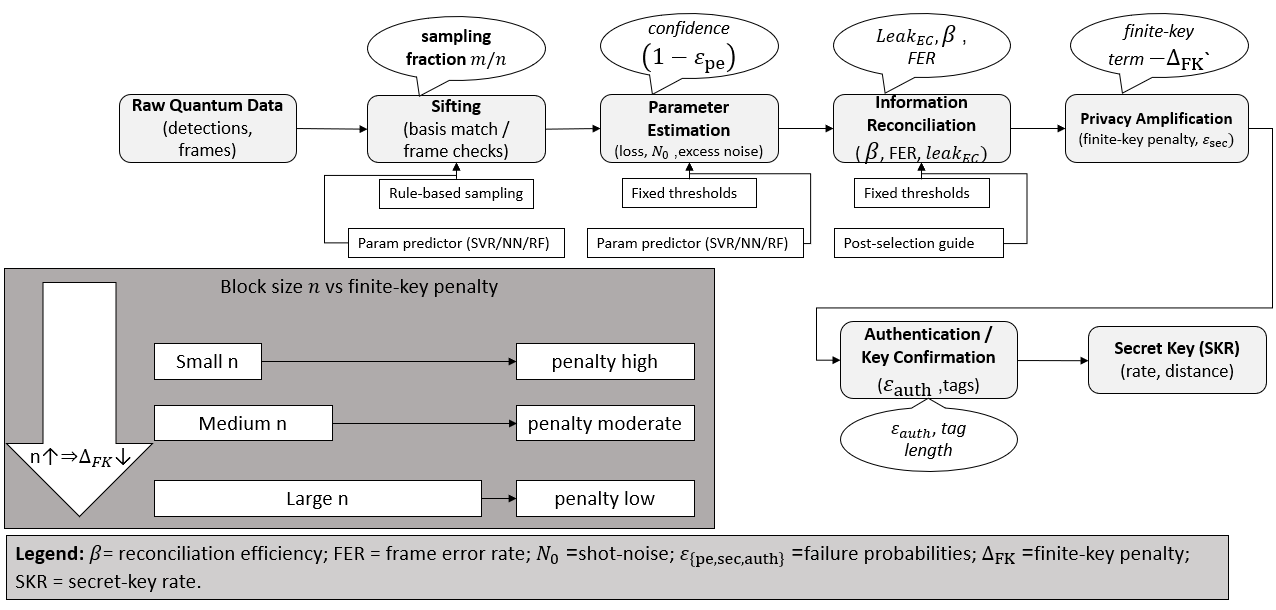}
  \caption{Protocol pipeline with screening hooks and finite\mbox{-}key coupling (larger $n$ $\rightarrow$ smaller penalty).}
  \label{fig:protocol-pipeline}
\end{figure}

Figure~\ref{fig:protocol-pipeline} sketches the left-to-right protocol flow—sifting, parameter estimation, reconciliation, authentication—with tap points for classical rules versus ML screens, and notes that increasing block size n reduces the finite-key penalty.

\textbf{Classical solutions.}
Longer blocks and tighter thresholds (with finite-key penalties); multidimensional reconciliation and Bayesian inference to improve operating points~\cite{num03_Leverrier2008_MultidimensionalReconciliationCVQKD,num86_Kleis2019_Improving_SKR_BayesianInference_CoherentQKD}. (Table~\ref{tab:master_VA_to_VF}, row \textbf{P4}, “Classical”.)

\textbf{ML solutions.}
Regression and supervised learning to predict optimal parameters and track drifts in real time ~\cite{num72_Liu2017_Monitoring_CVQKD_RealEnvironment,num21_Liu2018_ML_AutomaticParameterPrediction_CVQKD,num23_Wang2019_ML_OptimalParameterPrediction_QKD,num24_Ding2020_RandomForest_ParamPrediction_QKD, num25_Su2019_BP_NN_ParameterOptimization_Atmosphere_CVQKD,num41_Liu2019_PhaseModulationStabilization_ML_QKD, mohamed2026optiqkd}. (Table~\ref{tab:master_VA_to_VF}, row \textbf{P4}, “ML”.)

\textbf{Metaheuristic parameter optimization with integrated ML (CV)}
Rajendran \textit{et al.} propose a CV-QKD pipeline that couples metaheuristic parameter optimization with in-loop ML monitors \cite{num114_Rajendran2025_CVQKD_2EHRIO_ADS_CNN}. The method uses an Elitist Elk Herd Random Immigrants Optimizer (2E-HRIO) to tune system parameters (aiming to mitigate decoherence), alongside (i) a failure-prediction stage based on a radial DBSCAN-style classifier, and (ii) an ADS-CNN for side-channel attack identification; the authors also report reconciliation efficiency above $90\%$ and an increased key generation rate. Within our template, this primarily targets the \emph{protocol/process} layer (parameter choice, reconciliation operating point), while touching the device/measurement layer via defect/attack classification. Open questions include: coupling the optimizer’s decisions to composable finite-key bounds; reporting $\Delta$SKR and max distance at matched operating points; and inference/optimization latency on realistic controllers. (Table~\ref{tab:master_VA_to_VF}, row \textbf{P4-M}.)

\textbf{Comparative analysis.}
The protocol-layer comparison (Table~\ref{tab:master_VA_to_VF}, rows \textbf{P4} and \textbf{P4-M}) puts a structural tension on display. The classical lever for raising SKR is longer blocks plus multidimensional or Bayesian reconciliation: it works, it is auditable, and its security cost—the finite-key penalty as a function of block size—is fully characterized~\cite{num03_Leverrier2008_MultidimensionalReconciliationCVQKD,num86_Kleis2019_Improving_SKR_BayesianInference_CoherentQKD,num74_Furrer2012_FiniteKey_ComposableSecurity_CoherentAttacks_PRL,num75_Leverrier2017_Security_CVQKD_GaussianDeFinetti}. The ML lever is regression-based parameter prediction that tracks drift on a timescale faster than block-level reconfiguration~\cite{num21_Liu2018_ML_AutomaticParameterPrediction_CVQKD,num23_Wang2019_ML_OptimalParameterPrediction_QKD,num24_Ding2020_RandomForest_ParamPrediction_QKD,num25_Su2019_BP_NN_ParameterOptimization_Atmosphere_CVQKD,num72_Liu2017_Monitoring_CVQKD_RealEnvironment}, and recent metaheuristic+ML hybrids (2E-HRIO + radial-DBSCAN + ADS-CNN) report reconciliation efficiency above 90\% with a claimed key-generation-rate gain~\cite{num114_Rajendran2025_CVQKD_2EHRIO_ADS_CNN}. The two levers act on different terms in the SKR expression: the classical one improves the asymptotic $\beta$ and the finite-key correction, the ML one improves the operating point that those expressions are evaluated \emph{at}. They should compose multiplicatively rather than competitively, but composing them rigorously requires exposing the ML predictor's confidence to the finite-key calculator, which is precisely the certified-coupling gap the corpus has not yet closed.

\textbf{Open gaps.}
Certified coupling of ML predictors with finite-key bounds; reporting standards for latency/compute overheads; generalization across hardware.

\subsubsection*{P5. Sifting anomalies \& high-throughput screening}
\textbf{Problem \& evidence.}
Short-sample sifting schemes aim to flag disturbed signals quickly while consuming only a small fraction of raw keys~\cite{num98_Wu2022_SiftingScheme_CVQKD_ShortSamples_JOSAB}; process-level artifacts may leak patterns if not monitored \cite{num96_Scarani2009_Security_PracticalQKD_RMP,num97_Fung2007_PhaseRemappingAttack_PRA}. See Table~\ref{tab:master_VA_to_VF}, row \textbf{P5}.

\textbf{Classical solutions.}
Fixed sampling windows and rule-based screens integrated into sifting. (Table~\ref{tab:master_VA_to_VF}, row \textbf{P5}, “Classical”.)

\textbf{ML solutions.}
Fast anomaly screens during sifting (lightweight classifiers) that operate on sub-block summaries, with thresholds tuned to bound false alarms~\cite{num30_Mao2020_ML_DetectingQuantumAttacks_CVQKD,num31_Du2022_MultiAttackDetection_NN_CVQKD, num37_Liao2022_DBSCAN_DetectingAttacks_CVQKD,num98_Wu2022_SiftingScheme_CVQKD_ShortSamples_JOSAB}. (Table~\ref{tab:master_VA_to_VF}, row \textbf{P5}, “ML”.)

\textbf{Open gaps.}
Characterizing the privacy leakage vs.\ screening rate; integration with authentication budgets; dataset design for rare-event detection.

\textbf{Quantum-native ML for protocol optimization (DV)}
Beyond classical ML predictors and anomaly screens, \emph{quantum} machine learning has been explored to optimize protocol decision logic for DV-QKD. In \cite{QNNQRL2025}, QNN and QRL are integrated with BB84/B92, yielding QNN-BB84, QNN-B92, and QNN-QRL variants (QNN-QRL-V.1/V.2). The study reports gains on supervised metrics (Accuracy/Precision/Recall/F1, confusion matrices, ROC) and tests robustness across noisy channels. Within our template, these methods target the \emph{protocol/process} layer (decision policies) and are complementary to classical parameter tuning: potential benefits include improved key-generation quality under noise; open questions include (i) composable finite-key coupling of learned decisions, (ii) runtime/latency on realistic controllers, and (iii) generalization across hardware and channels. (Table~\ref{tab:master_VA_to_VF}, row \textbf{P4-Q}.)

\subsection{ML-Layer Vulnerabilities}
\subsubsection*{P6. Adversarial examples against QKD attack detectors}
\begin{figure}[!t]
  \centering
  \includegraphics[width=\columnwidth]{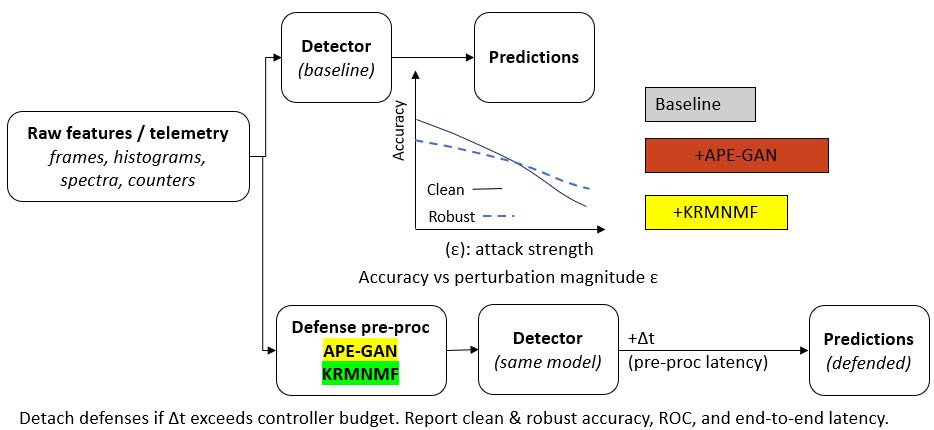}
  \caption{Adversarial robustness pipeline for QKD attack detectors and schematic clean–robust trade-offs (accuracy vs.\ perturbation, added latency).}
  \label{fig:adv-robustness}
\end{figure}

\textbf{Problem \& evidence.}
ML-based detectors can be degraded by carefully crafted perturbations; CV-QKD-specific demonstrations include one-pixel attacks and learned perturbations~\cite{num32_Tang2023_APE_GAN_Defense_CVQKD,num27_Guo2023_OnePixelAttack_CVQKD,num61_Lu2020_QuantumAdversarialML_PRResearch,al2026resisting}.  Broader QML threat taxonomies in comms systems echo these concerns and survey defenses such as adversarial training and data sanitization~\cite{num115_Nguyen_6G_QML_AdversarialThreats_IEEE}. See Table~\ref{tab:master_VA_to_VF}, row \textbf{P6}.
Fig.~\ref{fig:adv-robustness} contrasts the baseline detector path with defense pre-processing (APE-GAN/KRMNMF), and sketches the clean–robust accuracy trade-off and added latency that must fit the controller budget.

\textbf{Classical solutions.}
Model-agnostic pre-filters and randomized smoothing to dampen small adversarial perturbations. (Table~\ref{tab:master_VA_to_VF}, row \textbf{P6}, “Classical”.)

\textbf{ML solutions.}
Adversarial training; purification via APE-GAN and robust manifold reconstructions (e.g., KRMNMF) tailored to CV-QKD; evaluation includes clean vs.\ robust accuracy trade-offs and runtime~\cite{num32_Tang2023_APE_GAN_Defense_CVQKD, num58_Fu2023_KRMNMF_Defense_CVQKD,num61_Lu2020_QuantumAdversarialML_PRResearch}. (Table~\ref{tab:master_VA_to_VF}, row \textbf{P6}, “ML”.)

\textbf{Comparative analysis.}
The P6 row (Table~\ref{tab:master_VA_to_VF}) reports robust accuracies of 74.88\% (APE-GAN), 71.6\% (KRMNMF), and 78.8–79.5\% for the combination of the two~\cite{num32_Tang2023_APE_GAN_Defense_CVQKD,num58_Fu2023_KRMNMF_Defense_CVQKD,num27_Guo2023_OnePixelAttack_CVQKD,num61_Lu2020_QuantumAdversarialML_PRResearch}. Three readings of these numbers matter for deployment. \emph{First}, no single defense exceeds 80\% robust accuracy in the surveyed regime, which sets a hard ceiling on what an inline purification stage can deliver and means adversarial robustness must always be paired with conservative finite-key fallbacks—an ML detector that is right 4 out of 5 times under perturbation cannot be a sole gating function on key acceptance. \emph{Second}, the combination's $\le$5 percentage-point gain over the better individual defense is consistent with the two methods removing partially overlapping rather than orthogonal perturbations, suggesting that further additive defenses will face diminishing returns and that diversity of defense \emph{mechanism}, not just stacking, is what the next generation needs. \emph{Third}, the corpus reports robust accuracy but rarely reports the induced controller-cycle latency at the same operating point, and inline purification only earns its keep when its latency fits inside the existing controller budget; the missing measurement is the same one we flagged for P3 and P4.

\textbf{Open gaps.}
Standardized adversarial test suites for QKD telemetry; certified robustness bounds; defenses that preserve SKR under finite-key constraints.

\subsection{Network-Level Security (IDS, Coexistence, Resource Allocation)}
\subsubsection*{P7. Coexistence with classical traffic (SWDM/SDM) and optical impairments}
\begin{figure}[!t]
  \centering
  \includegraphics[width=\columnwidth]{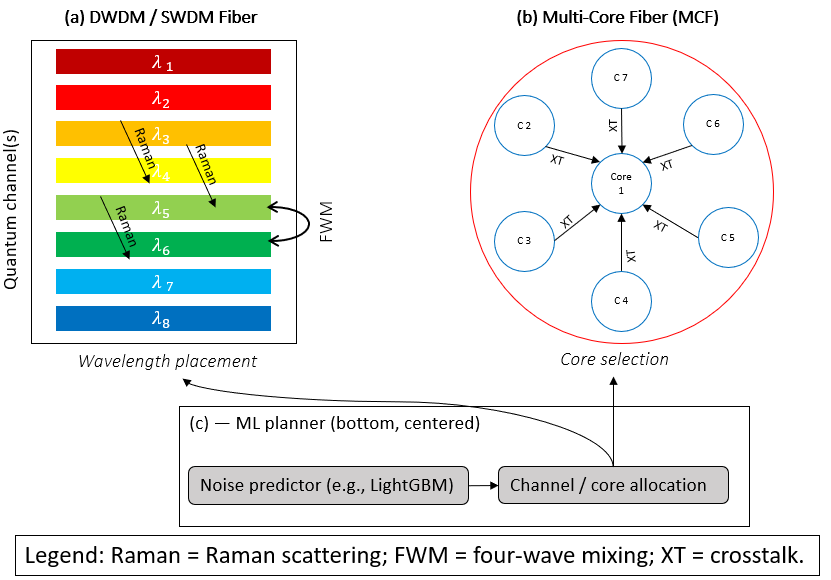}
  \caption{Coexistence impairments (Raman, FWM, inter-core XT) and ML-guided wavelength/core allocation in DWDM/MCF links.}
  \label{fig:coexistence}
\end{figure}

\textbf{Problem \& evidence.}
Raman scattering, four-wave mixing, and inter-core crosstalk in co-propagation settings degrade quantum channels~\cite{num78_Wang2022_NoisePrediction_ML_Secured_SWDM_B5G_Fronthaul,num79_Markowski2016_FWM_DWDM_1310nm_ApplOpt,num80_Kong2022_CoreWavelengthAllocation_SNS_QKD_MCF_OFC, num81_Niu2019_LightGBM_NoiseSuppressing_DWDM_QKD,num83_Winzer2018_FiberOptic_Transmission_Networking_20Years, num84_Napoli2018_TowardsMultibandOpticalSystems_PND}; studies report noise prediction and channel/core allocation strategies (including ML)~\cite{num78_Wang2022_NoisePrediction_ML_Secured_SWDM_B5G_Fronthaul, num81_Niu2019_LightGBM_NoiseSuppressing_DWDM_QKD}. See Table~\ref{tab:master_VA_to_VF}, row \textbf{P7}. Figure \ref{fig:coexistence} summarizes coexistence impairments in DWDM/MCF (Raman, FWM, inter-core XT) and shows how a LightGBM-style noise predictor can drive wavelength/core allocation.

\textbf{Classical solutions.}
Static provisioning and guard-band/channel planning; impairment-aware allocation~\cite{num83_Winzer2018_FiberOptic_Transmission_Networking_20Years,num79_Markowski2016_FWM_DWDM_1310nm_ApplOpt,num84_Napoli2018_TowardsMultibandOpticalSystems_PND}. (Table~\ref{tab:master_VA_to_VF}, row \textbf{P7}, “Classical”.)

\textbf{ML solutions.}
Noise prediction (e.g., LightGBM) and data-driven allocation policies to reduce evaluation time and improve placement~\cite{num78_Wang2022_NoisePrediction_ML_Secured_SWDM_B5G_Fronthaul, num81_Niu2019_LightGBM_NoiseSuppressing_DWDM_QKD}; learning-based optimization for multiband/MCF contexts. (Table~\ref{tab:master_VA_to_VF}, row \textbf{P7}, “ML”.)

\textbf{Open gaps.}
Joint optimization of optical-layer coexistence with SKR targets; transferability across network topologies.

\subsubsection*{P8. IDS integration and operational security}
\textbf{Problem \& evidence.}
Classical IDS faces signature brittleness and anomaly false positives; integrating IDS around QKD controllers and interfaces augments operational security~\cite{num99_Alahmadi2022_CyberSecurity_Threats_SideChannel_DigitalAgriculture_Sensors,num100_Chkirbene2020_TIDCS_DynamicIDS_FeatureSelection_Access,num101_Hijazi2018_DL_IDS_IndustryNetwork_BDCSIntell,num102_Vacca2012_ComputerInformationSecurityHandbook,num103_Liao2013_IDS_ComprehensiveReview_JNCA,num104_Tama2019_TSEIDS_TwoStageClassifier_AnomalyIDS_Access,num105_GarciaTeodoro2009_AnomalyBased_IDS_Survey_CompSec,num106_Gumusbas2021_Survey_Databases_DL_Cybersecurity_IDS_IEEE_SystemsJ,num109_Sirisha2023_EfficientMLTechniques_IDS_ICSCDS,num110_Srilakshmi2022_SecureOptimizationRouting_MANETs_Access,num111_Hung2020_EnergyEfficient_CooperativeRouting_Heterogeneous_WSN_Access}. At the vehicular-network layer, blockchain-based event validation and authentication can complement QKD by strengthening identity, trust, and data-integrity services outside the quantum link itself~\cite{al2020reputation}. See Table~\ref{tab:master_VA_to_VF}, row \textbf{P8}.

\textbf{Classical solutions.}
Signature and anomaly-based IDS pipelines; curated feature engineering; ensemble methods~\cite{num100_Chkirbene2020_TIDCS_DynamicIDS_FeatureSelection_Access,num104_Tama2019_TSEIDS_TwoStageClassifier_AnomalyIDS_Access}. (Table~\ref{tab:master_VA_to_VF}, row \textbf{P8}, “Classical”.)

\textbf{ML solutions.}
Deep and ensemble IDS tailored to traffic/telemetry distributions in relevant deployments (industrial, IoT/railway scenarios)~\cite{num70_AlMohammed2021_ML_DetectAttackers_QKD_IoT_Railway_IEEEAccess,num101_Hijazi2018_DL_IDS_IndustryNetwork_BDCSIntell}. (Table~\ref{tab:master_VA_to_VF}, row \textbf{P8}, “ML”.)

\textbf{Open gaps.}
Dataset realism and drift handling; privacy-preserving sharing of traces; aligning IDS alerts with QKD-layer responses.

\subsubsection*{P9. Resource allocation in QKD networks}
\textbf{Problem \& evidence.}
Multi-tenant provisioning and trusted/untrusted relay design on backbones/metropolitan settings impose scheduling and trust trade-offs~\cite{num01_Cao2020_MultiTenantProvisioningQKDNet,num95_Cao2021_HybridTrustedUntrustedRelay_QKD_Backbone_JSAC}; privacy-preserving edge learning for resource control is also explored~\cite{num94_Xu2023_PrivacyPreserving_ResourceAllocation_FederatedEdge_QI_JSTSP}, high altitude platform-based XOR-relay QKD extends trusted-relay design to non-terrestrial 6G settings and is therefore relevant when comparing relay architectures, hardened links, and finite-key-aware deployment strategies~\cite{al2025quantum}. See Table~\ref{tab:master_VA_to_VF}, row \textbf{P9}.

\textbf{Classical solutions.}
Heuristics and graph-based provisioning with trust constraints; static or periodically re-optimized allocations~\cite{num01_Cao2020_MultiTenantProvisioningQKDNet,num95_Cao2021_HybridTrustedUntrustedRelay_QKD_Backbone_JSAC}. (Table~\ref{tab:master_VA_to_VF}, row \textbf{P9}, “Classical”.)

\textbf{ML solutions.}
Learning-assisted policies (including RL variants) for dynamic demand and uncertainty; federated learning for privacy preservation at the edge~\cite{num94_Xu2023_PrivacyPreserving_ResourceAllocation_FederatedEdge_QI_JSTSP,num115_Nguyen_6G_QML_AdversarialThreats_IEEE}; evaluation by throughput/SKR fairness and reconfiguration latency. (Table~\ref{tab:master_VA_to_VF}, row \textbf{P9}, “ML”.)

\textbf{Comparative analysis.}
The P9 row sits at a different scale from P1–P6: the unit of decision is no longer a block or a shot but a tenant, a wavelength, or a relay. Classical heuristics and graph-based provisioning are highly interpretable but tend to underutilize capacity in the face of bursty demand and uncertain SKR estimates~\cite{num01_Cao2020_MultiTenantProvisioningQKDNet,num95_Cao2021_HybridTrustedUntrustedRelay_QKD_Backbone_JSAC}. Learning-assisted, RL, and federated edge policies trade some interpretability for adaptivity~\cite{num94_Xu2023_PrivacyPreserving_ResourceAllocation_FederatedEdge_QI_JSTSP,num115_Nguyen_6G_QML_AdversarialThreats_IEEE} and may, in principle, deliver higher throughput-fairness on real traces. The key open variable, however, is robustness to traffic bursts and failure modes—precisely the regime where classical heuristics' worst-case behavior is best understood—so we view ML at this layer as most credible when used \emph{offline} for planning and what-if analysis, with the live policy held to a small, auditable rule set that the ML planner has informed but not replaced.

\textbf{Open gaps.}
Online evaluation on real traces; robustness to traffic bursts and failures; interfaces between network policies and link-layer SKR estimators.

\section{Benchmarking \& Evaluation Framework}
\label{sec:benchmarking}

The case for a compact, TETCI-friendly benchmarking framework is empirical, not aesthetic: the comparative analyses in Section~\ref{sec:problem-centric} repeatedly identified the same missing measurement—matched-operating-point $\Delta$SKR, distance, and latency—across P3, P4, P6, and V-C. Until these are reported at common operating points, the corpus' headline accuracy numbers (DBSCAN F1 of 0.998, robust accuracies in the 70–80\% range, decision-tree CA detection of 100\%/91.26\%) cannot be combined into a deployment recommendation. Simulation-based QKD studies can support reproducible evaluation environments for protocol behavior, attack modeling, and defense testing before hardware deployment~\cite{al2021new}; what follows is a grounded proposal for what such a framework should look like, anchored exclusively in the works present in the corpus.

\subsection{Datasets and feature suites}
\textbf{Data sources.} We use (i) \emph{lab/field telemetry} from CV-QKD monitoring and long-distance trials~\cite{num11_Huang2016_LongDistanceCVQKD_ExcessNoiseControl,num12_Jouguet2013_LongDistanceCVQKD_ExperimentalDemo,num72_Liu2017_Monitoring_CVQKD_RealEnvironment,num02_Hajomer2022_CVQKD60kmRLO}; (ii) \emph{coexistence traces} from SWDM/SDM studies~\cite{num78_Wang2022_NoisePrediction_ML_Secured_SWDM_B5G_Fronthaul,num79_Markowski2016_FWM_DWDM_1310nm_ApplOpt,num80_Kong2022_CoreWavelengthAllocation_SNS_QKD_MCF_OFC, num81_Niu2019_LightGBM_NoiseSuppressing_DWDM_QKD,num83_Winzer2018_FiberOptic_Transmission_Networking_20Years, num84_Napoli2018_TowardsMultibandOpticalSystems_PND,num85_Diamanti2015_DistributingSecretKeys_CV_QKD_Entropy}; and (iii) \emph{protocol/process logs} for sifting and parameter estimation~\cite{num98_Wu2022_SiftingScheme_CVQKD_ShortSamples_JOSAB,num47_Jouguet2012_Imperfections_Practical_CVQKD,num71_Wang2016_FiniteSamplingBandwidth_CVQKD}.
Ultra-high-speed and evacuated-tube train scenarios provide an extreme mobility case for evaluating beam alignment, outage behavior, and free-space channel assumptions relevant to future QKD planning~\cite{al2022free}. 
\textbf{Feature families.}
\begin{itemize}
  \item \textit{Device/measurement (CV):} LO power/current \(I_{\mathrm{LO}}\), shot-noise level \(N_{0}\), homodyne statistics (mean/variance/kurtosis), wavelength/detector-linearity indicators, saturation flags~\cite{num05_Ma2013_LOFluctuationLoophole_CVQKD,num06_Huang2013_WavelengthAttack_CVQKD,num07_Ma2013_WavelengthAttack_Heterodyne_CVQKD, num08_Jouguet2013_PreventingCalibrationAttacks_LO_CVQKD,num09_Qin2016_SaturationAttack_CVQKD,num10_Qin2018_HomodyneBlinding_CVQKD,num11_Huang2016_LongDistanceCVQKD_ExcessNoiseControl,num12_Jouguet2013_LongDistanceCVQKD_ExperimentalDemo,num72_Liu2017_Monitoring_CVQKD_RealEnvironment}.
  \item \textit{DV parallels:} timing histograms, after-gate windows, detector count-rate/illumination proxies \cite{num14_Lydersen2010_HackingCommercialQKD_BrightIllumination,num48_Wiechers2011_AfterGateAttack_NJP,num50_Qi2007_TimeShiftAttack_QIC,num97_Fung2007_PhaseRemappingAttack_PRA}.
  \item \textit{Channel/coexistence:} channel power, launch power per band/core, Raman/FWM proxies, inter-core crosstalk indices, impairment margins (from optical-layer studies)~\cite{num78_Wang2022_NoisePrediction_ML_Secured_SWDM_B5G_Fronthaul,num79_Markowski2016_FWM_DWDM_1310nm_ApplOpt,num80_Kong2022_CoreWavelengthAllocation_SNS_QKD_MCF_OFC, num81_Niu2019_LightGBM_NoiseSuppressing_DWDM_QKD,num83_Winzer2018_FiberOptic_Transmission_Networking_20Years, num84_Napoli2018_TowardsMultibandOpticalSystems_PND}.
  \item \textit{Protocol/process:} short-sample block statistics for sifting, reconciliation logs (iterations, \(\beta\)), authentication budget counters~\cite{num03_Leverrier2008_MultidimensionalReconciliationCVQKD,num86_Kleis2019_Improving_SKR_BayesianInference_CoherentQKD}, \cite{num96_Scarani2009_Security_PracticalQKD_RMP,num98_Wu2022_SiftingScheme_CVQKD_ShortSamples_JOSAB}.
\end{itemize}
\textbf{Modalities.} Label each dataset as \emph{fiber} vs.\ \emph{free-space} (generic stability context) \cite{num13_Xu2020_SecureQKDRealisticDevices}, and as \emph{public/simulated} vs.\ \emph{private/field}.

\subsection{Stress protocols}
\textbf{Adversarial suites.} Evaluate attack detectors under: (i) \emph{one-pixel/structured} perturbations~\cite{num27_Guo2023_OnePixelAttack_CVQKD}; (ii) \emph{purification-aware} attacks (to test APE-GAN/KRMNMF defenses)~\cite{num32_Tang2023_APE_GAN_Defense_CVQKD, num58_Fu2023_KRMNMF_Defense_CVQKD}; and (iii) \emph{standard gradient-based} attacks (e.g., FGSM/PGD/DeepFool; reported as generic adversarial baselines).
\textbf{Distribution shift.} Hardware swaps (detector, laser), environment/load changes (SWDM/SDM occupancy)~\cite{num78_Wang2022_NoisePrediction_ML_Secured_SWDM_B5G_Fronthaul,num81_Niu2019_LightGBM_NoiseSuppressing_DWDM_QKD}, operating-point drift (power/SNR)~\cite{num11_Huang2016_LongDistanceCVQKD_ExcessNoiseControl,num12_Jouguet2013_LongDistanceCVQKD_ExperimentalDemo}.
\textbf{Finite-key small-block regime.} Repeat evaluations at progressively smaller block sizes and report the induced finite-key penalties alongside detection outcomes~\cite{num96_Scarani2009_Security_PracticalQKD_RMP,num74_Furrer2012_FiniteKey_ComposableSecurity_CoherentAttacks_PRL,num75_Leverrier2017_Security_CVQKD_GaussianDeFinetti}.

\subsection{Metrics}
\textbf{Detection.} Precision/Recall/F1 (and ROC-type summaries when reported) for attack/anomaly detectors~\cite{num30_Mao2020_ML_DetectingQuantumAttacks_CVQKD,num31_Du2022_MultiAttackDetection_NN_CVQKD, num37_Liao2022_DBSCAN_DetectingAttacks_CVQKD}.
\textbf{Security impact.} \(\Delta\)SKR (with sign), and/or \emph{maximum distance} at fixed security level; report identical operating points for head-to-head comparisons~\cite{num11_Huang2016_LongDistanceCVQKD_ExcessNoiseControl,num12_Jouguet2013_LongDistanceCVQKD_ExperimentalDemo,num86_Kleis2019_Improving_SKR_BayesianInference_CoherentQKD}. Some studies report classifier accuracy and reconciliation efficiency (e.g., ADS-CNN with $\geq$90\% reconciliation efficiency reported in \cite{num114_Rajendran2025_CVQKD_2EHRIO_ADS_CNN}); for comparability we urge concurrent reporting of $\Delta$SKR and maximum distance at matched operating points, as well as optimizer/ML latency.

\textbf{Operational.} Inference latency (per block/shot), controller-loop compatibility, compute footprint (CPU/GPU/FPGA notes when available), and energy (if reported).
\textbf{Robustness.} Empirical robust accuracy under adversarial suites and shift; certified bounds if provided. For purification defenses (APE-GAN, KRMNMF), include pre/post metrics and induced latency~\cite{num32_Tang2023_APE_GAN_Defense_CVQKD, num58_Fu2023_KRMNMF_Defense_CVQKD,num61_Lu2020_QuantumAdversarialML_PRResearch}.
\textbf{Compliance cues.} Note any explicit alignment with typical security/operations practices summarized in surveys/roadmaps \cite{num13_Xu2020_SecureQKDRealisticDevices}, \cite{num96_Scarani2009_Security_PracticalQKD_RMP} (e.g., ETSI/ITU--T mentions where present in the source).

\subsection{Baselines}
\textbf{Classical filters.} LO power monitors, wavelength filtering, calibration hardening, detector linearity checks, and MDI/DI shifts where applicable~\cite{num11_Huang2016_LongDistanceCVQKD_ExcessNoiseControl,num12_Jouguet2013_LongDistanceCVQKD_ExperimentalDemo,num08_Jouguet2013_PreventingCalibrationAttacks_LO_CVQKD,num16_Acin2007_DeviceIndependentSecurity_CollectiveAttacks,num17_Lo2012_MDIQKD_PRL,num28_Tang2014_MDIQKD_Polarization_Experimental}.
\textbf{Minimal ML.} A small NN and a tree-based model (e.g., RF/LightGBM) as lightweight baselines for detection/forecasting~\cite{num21_Liu2018_ML_AutomaticParameterPrediction_CVQKD,num23_Wang2019_ML_OptimalParameterPrediction_QKD,num24_Ding2020_RandomForest_ParamPrediction_QKD, num25_Su2019_BP_NN_ParameterOptimization_Atmosphere_CVQKD,num30_Mao2020_ML_DetectingQuantumAttacks_CVQKD,num31_Du2022_MultiAttackDetection_NN_CVQKD, num37_Liao2022_DBSCAN_DetectingAttacks_CVQKD,num78_Wang2022_NoisePrediction_ML_Secured_SWDM_B5G_Fronthaul,num81_Niu2019_LightGBM_NoiseSuppressing_DWDM_QKD}.
\textbf{Purification baselines.} APE-GAN and robust-manifold reconstruction (e.g., KRMNMF) as pre-processing defenses for CV-QKD detectors~\cite{num32_Tang2023_APE_GAN_Defense_CVQKD, num58_Fu2023_KRMNMF_Defense_CVQKD}.
\textbf{Unsupervised.} Density-based clustering (DBSCAN) as a modality-agnostic detector~\cite{num37_Liao2022_DBSCAN_DetectingAttacks_CVQKD}.

\subsection{Reproducibility checklist}
For every experiment, report:
\begin{itemize}
  \item \textbf{Data:} source (sim/lab/field), license/access, feature definitions, normalization, and exact train/val/test splits; handling of class imbalance~\cite{num72_Liu2017_Monitoring_CVQKD_RealEnvironment,num30_Mao2020_ML_DetectingQuantumAttacks_CVQKD,num31_Du2022_MultiAttackDetection_NN_CVQKD, num37_Liao2022_DBSCAN_DetectingAttacks_CVQKD,num98_Wu2022_SiftingScheme_CVQKD_ShortSamples_JOSAB}.
  \item \textbf{Code \& config:} repository link (if available), commit hash, seed control, hyperparameter search protocol, early-stopping criteria.
  \item \textbf{Compute:} hardware (CPU/GPU/FPGA), batch size, wall-clock; online vs.\ offline path and controller-rate compatibility.
  \item \textbf{Stress protocol spec:} attack parameters (for adversarial tests), shift scenarios (hardware/load), block sizes for finite-key sweeps~\cite{num32_Tang2023_APE_GAN_Defense_CVQKD, num58_Fu2023_KRMNMF_Defense_CVQKD,num74_Furrer2012_FiniteKey_ComposableSecurity_CoherentAttacks_PRL,num75_Leverrier2017_Security_CVQKD_GaussianDeFinetti,num27_Guo2023_OnePixelAttack_CVQKD}.
  \item \textbf{Reporting:} detection (P/R/F1), \(\Delta\)SKR and/or distance at matched operating points, latency/compute, and robustness; note any standards/compliance statements present in the original paper \cite{num13_Xu2020_SecureQKDRealisticDevices}, \cite{num96_Scarani2009_Security_PracticalQKD_RMP}.
\end{itemize}

{
\refstepcounter{table}\label{tab:eval-grid}
\noindent\textbf{Table \thetable.} Evaluation grid linking datasets, stress protocols, and required metrics.

\par\noindent
\begin{minipage}{\columnwidth}
  \centering
  \scriptsize
  \setlength{\tabcolsep}{1.5pt}
  \renewcommand{\arraystretch}{1.03}

  \newcommand{\mstack}{%
    \begingroup\tiny\renewcommand{\arraystretch}{0.9}%
      \begin{tabular}[c]{@{}c@{}}%
        P/R/F1\\
        $\Delta$SKR\\
        Max distance\\
        Latency\\
        Robustness%
      \end{tabular}%
    \endgroup
  }

  \begin{tabular}{@{}p{0.25\columnwidth} p{0.19\columnwidth} p{0.19\columnwidth} p{0.19\columnwidth} p{0.15\columnwidth}@{}}
    \hline
    \textbf{Dataset} & \textbf{Adv. (adversarial)} & \textbf{Shift (distribution shift)} & \textbf{Small-block (finite-key)} & \textbf{Notes} \\
    \hline
    Fiber — Lab        & \mstack & \mstack & \mstack & bench; repeatable; high rate \\
    Fiber — Field      & \mstack & \mstack & \mstack & live impairments; drift \\
    Free-space — Lab   & \mstack & \mstack & \mstack & turbulence emulator; alignment \\
    Free-space — Field & \mstack & \mstack & \mstack & weather; pointing; outages \\
    \hline
  \end{tabular}

  \vspace{0.25em}
  \scriptsize P/R/F1 = Precision/Recall/F1; $\Delta$SKR = change in secret-key rate.
\end{minipage}
}

Table~\ref{tab:eval-grid} summarizes the evaluation grid (Datasets × Stress × Metrics) we use across case studies in Secs.~V–VI


\section{Systems Integration \& Deployment Lessons}
\label{sec:systems}

The transition from a published defense to a deployed one is governed by a small number of practical decisions—modality and LO architecture, monitor placement and insertion-loss budget, controller latency and post-processing scheduling, and which ML components are actually \emph{shippable} today versus research-stage. This section distills the corpus along those four axes, with the explicit goal of separating ``works in the lab'' from ``ready to deploy.''

\subsection{CV vs.\ DV; LLO vs.\ TLO (Architecture Choices)}
\textbf{Modality.} DV and CV offer complementary operating regimes and implementation risks, as surveyed in~\cite{num04_Gisin2002_QuantumCryptographyReview, num29_Li2017_CVQKD_ChinesePhysicsB_Review, num96_Scarani2009_Security_PracticalQKD_RMP,num13_Xu2020_SecureQKDRealisticDevices}. 
\textbf{LO handling.} For CV, the choice between transmitted-LO (TLO) and local-LO (LLO) impacts both practicality and attack surface. Field/long-distance demonstrations and real-environment monitoring indicate viable operating points under LLO/TLO configurations~\cite{num11_Huang2016_LongDistanceCVQKD_ExcessNoiseControl,num12_Jouguet2013_LongDistanceCVQKD_ExperimentalDemo,num72_Liu2017_Monitoring_CVQKD_RealEnvironment}, with an LLO-based 60\,km fiber result illustrating complexity vs.\ performance trade-offs~\cite{num02_Hajomer2022_CVQKD60kmRLO}. 
\textbf{Security implications.} LO-related side-channels (fluctuation, wavelength, calibration) remain primary concerns in CV and motivate hardening or architectural shifts (e.g., MDI/DI) ~\cite{num05_Ma2013_LOFluctuationLoophole_CVQKD,num06_Huang2013_WavelengthAttack_CVQKD,num07_Ma2013_WavelengthAttack_Heterodyne_CVQKD, num08_Jouguet2013_PreventingCalibrationAttacks_LO_CVQKD,num16_Acin2007_DeviceIndependentSecurity_CollectiveAttacks,num28_Tang2014_MDIQKD_Polarization_Experimental}. Coherent-state CV foundations appear in~\cite{num42_Grosshans2002_CVQKD_CoherentStates_PRL}.

\subsection{Chip-Based \& Implementation Concerns}
\textbf{Integrated threats.} Power-analysis style leakage on integrated CV-QKD systems has been demonstrated, underscoring board-/chip-level telemetry as an attack vector~\cite{num52_Zheng2021_PowerAnalysis_Integrated_CVQKD}. Optical components may also face physical stress (e.g., laser-damage on attenuators), calling for margins and monitoring in the opto-electronic chain~\cite{num55_Huang2020_LaserDamage_Attenuators_QKD_PRAppl}. 
\textbf{Controller budgets.} Practical control/processing stacks operate under latency/energy constraints; system-level techniques (e.g., per-core DVFS) inform compute provisioning for embedded controllers running estimation, reconciliation, or ML inference~\cite{num53_Kim2008_PerCoreDVFS_HPCA}. Classical countermeasures to side-channel analysis in electronics (e.g., on-chip suppression) highlight the need to consider cross-layer leakage paths~\cite{num51_Ratanpal2004_OnChipSignalSuppression_PowerAnalysis}.

\subsection{Co-Propagation with Classical Channels (SWDM/SDM/Multiband)}
\textbf{Optical impairments.} When quantum channels co-propagate with classical traffic, Raman scattering, four-wave mixing, and inter-core crosstalk can degrade estimation and SKR~\cite{num83_Winzer2018_FiberOptic_Transmission_Networking_20Years,num79_Markowski2016_FWM_DWDM_1310nm_ApplOpt,num84_Napoli2018_TowardsMultibandOpticalSystems_PND}. 
\textbf{Planning \& adaptation.} Studies propose noise prediction and placement policies for channel/core allocation (including LightGBM and related models), reducing evaluation time for dynamic DWDM-QKD networks and multicores~\cite{num78_Wang2022_NoisePrediction_ML_Secured_SWDM_B5G_Fronthaul,num81_Niu2019_LightGBM_NoiseSuppressing_DWDM_QKD,num80_Kong2022_CoreWavelengthAllocation_SNS_QKD_MCF_OFC}. Broader traffic and spectrum evolution motivate multiband-aware designs~\cite{num83_Winzer2018_FiberOptic_Transmission_Networking_20Years,num84_Napoli2018_TowardsMultibandOpticalSystems_PND}, with surveys on CV implementations providing the security context~\cite{num85_Diamanti2015_DistributingSecretKeys_CV_QKD_Entropy}.

\subsection{IoT Controllers, Edge Offloading, and Industrial Contexts}
\textbf{Edge/IoT use-cases.} QKD has been considered to harden IoT scenarios (including rail/industrial) with ML-based attacker detection during the quantum phase \cite{num46_AlMohammed2021_GLOBECOM_NN_QKD_IoT, num69_AlMohammed2021_OnUseQuantumComm_SecuringIoT_6G_ICCWS,num70_AlMohammed2021_ML_DetectAttackers_QKD_IoT_Railway_IEEEAccess}. 
\textbf{Compute placement.} Fog/edge paradigms guide where to terminate control loops and offload analytics \cite{num44_Dastjerdi2016_FogComputing_IoT_Potential}. For broader quantum networking, privacy-preserving/federated learning provides a template for distributing optimization at the edge while respecting data locality~\cite{num94_Xu2023_PrivacyPreserving_ResourceAllocation_FederatedEdge_QI_JSTSP}. Experimental demonstrations targeting energy-efficient SD-IoT highlight the systems direction of travel~\cite{num68_Mavromatis_Experimental_QKD_EnergyEfficient_SDIoT}. 
\textbf{Classical IDS interplay.} Integration with classical IDS (signature/anomaly/ensembles) around QKD controllers improves operational visibility, but inherits dataset/shift challenges \cite{ num99_Alahmadi2022_CyberSecurity_Threats_SideChannel_DigitalAgriculture_Sensors,num100_Chkirbene2020_TIDCS_DynamicIDS_FeatureSelection_Access,num101_Hijazi2018_DL_IDS_IndustryNetwork_BDCSIntell,num102_Vacca2012_ComputerInformationSecurityHandbook,num103_Liao2013_IDS_ComprehensiveReview_JNCA,num104_Tama2019_TSEIDS_TwoStageClassifier_AnomalyIDS_Access,num105_GarciaTeodoro2009_AnomalyBased_IDS_Survey_CompSec,num106_Gumusbas2021_Survey_Databases_DL_Cybersecurity_IDS_IEEE_SystemsJ,num109_Sirisha2023_EfficientMLTechniques_IDS_ICSCDS,num110_Srilakshmi2022_SecureOptimizationRouting_MANETs_Access,num111_Hung2020_EnergyEfficient_CooperativeRouting_Heterogeneous_WSN_Access}.  Recent 6G and metaverse-oriented surveys highlight semantic communication and edge learning as enabling technologies, suggesting that future QKD monitoring may need to interoperate with distributed intelligence and edge-native control planes~\cite{aloudat2025metaverse}.

\subsection{Authentication \& Post-Processing Budgets}
\textbf{Reconciliation/authentication.} Reconciliation efficiency and authentication costs influence effective SKR and latency; multidimensional/Bayesian improvements help tune operating points~\cite{num03_Leverrier2008_MultidimensionalReconciliationCVQKD,num86_Kleis2019_Improving_SKR_BayesianInference_CoherentQKD}. System-level reviews emphasize that authenticated classical exchanges are assumed and must be budgeted in deployments \cite{num13_Xu2020_SecureQKDRealisticDevices}, \cite{num96_Scarani2009_Security_PracticalQKD_RMP}. Finite-size and bandwidth-limited estimation further couple block sizing with delay and security margins~\cite{num74_Furrer2012_FiniteKey_ComposableSecurity_CoherentAttacks_PRL,num75_Leverrier2017_Security_CVQKD_GaussianDeFinetti,num47_Jouguet2012_Imperfections_Practical_CVQKD,num71_Wang2016_FiniteSamplingBandwidth_CVQKD}.

\subsection{What Actually Ships \& Where ML Is Ready}
\textbf{Fielded capabilities.} Long-distance and field/real-environment CV-QKD demonstrations~\cite{num11_Huang2016_LongDistanceCVQKD_ExcessNoiseControl,num12_Jouguet2013_LongDistanceCVQKD_ExperimentalDemo,num72_Liu2017_Monitoring_CVQKD_RealEnvironment}, alongside network-level provisioning studies for backbones/metro settings \cite{ num01_Cao2020_MultiTenantProvisioningQKDNet,num95_Cao2021_HybridTrustedUntrustedRelay_QKD_Backbone_JSAC}, indicate maturing engineering stacks (hardware hardening + calibrated post-processing).
\textbf{ML components near deployment.}
\begin{itemize}
  \item \emph{Parameter prediction \& stabilization/monitoring (CV):} multiple works report ANN/SVR/RF predictors and stabilization aids with lab/field telemetry~\cite{num72_Liu2017_Monitoring_CVQKD_RealEnvironment,num21_Liu2018_ML_AutomaticParameterPrediction_CVQKD,num23_Wang2019_ML_OptimalParameterPrediction_QKD,num24_Ding2020_RandomForest_ParamPrediction_QKD, num25_Su2019_BP_NN_ParameterOptimization_Atmosphere_CVQKD,num41_Liu2019_PhaseModulationStabilization_ML_QKD}. These are closest to near-term integration where controller cycles and features are well-defined.
  \item \emph{Attack/anomaly detectors:} supervised and density-based detectors (DBSCAN) show promising P/R/F1 in lab-style settings~\cite{num30_Mao2020_ML_DetectingQuantumAttacks_CVQKD,num31_Du2022_MultiAttackDetection_NN_CVQKD, num37_Liao2022_DBSCAN_DetectingAttacks_CVQKD}; productionization hinges on drift handling and adversarial stress evidence.
  \item \emph{Coexistence planners:} noise prediction and LightGBM-based allocation for DWDM/MCF are promising for NOC/offline planning and fast what-if evaluation~\cite{num78_Wang2022_NoisePrediction_ML_Secured_SWDM_B5G_Fronthaul,num81_Niu2019_LightGBM_NoiseSuppressing_DWDM_QKD}.
\end{itemize}
\textbf{Research-stage ML.} Adversarial purification/robustness (APE-GAN, manifold NMF) \cite{num32_Tang2023_APE_GAN_Defense_CVQKD, num58_Fu2023_KRMNMF_Defense_CVQKD} and broader adversarial learning~\cite{num61_Lu2020_QuantumAdversarialML_PRResearch,num59_Ren2022_ExperimentalQuantumAdversarialLearning, num60_Yan2023_ArtificialKeyFingerprints_CVQKD} remain active research areas; deploying these inline requires latency accounting, stability analysis, and standardized stress protocols.

Integrated stacks that combine parameter optimization with inline failure/attack screening (e.g., 2E-HRIO + radial DBSCAN + ADS-CNN) \cite{num114_Rajendran2025_CVQKD_2EHRIO_ADS_CNN} illustrate a trend toward joint protocol/process tuning and device monitoring; productionization requires HIL evaluation and controller-cycle budgeting.

\begin{figure}[!t]
  \centering
  \includegraphics[width=\columnwidth]{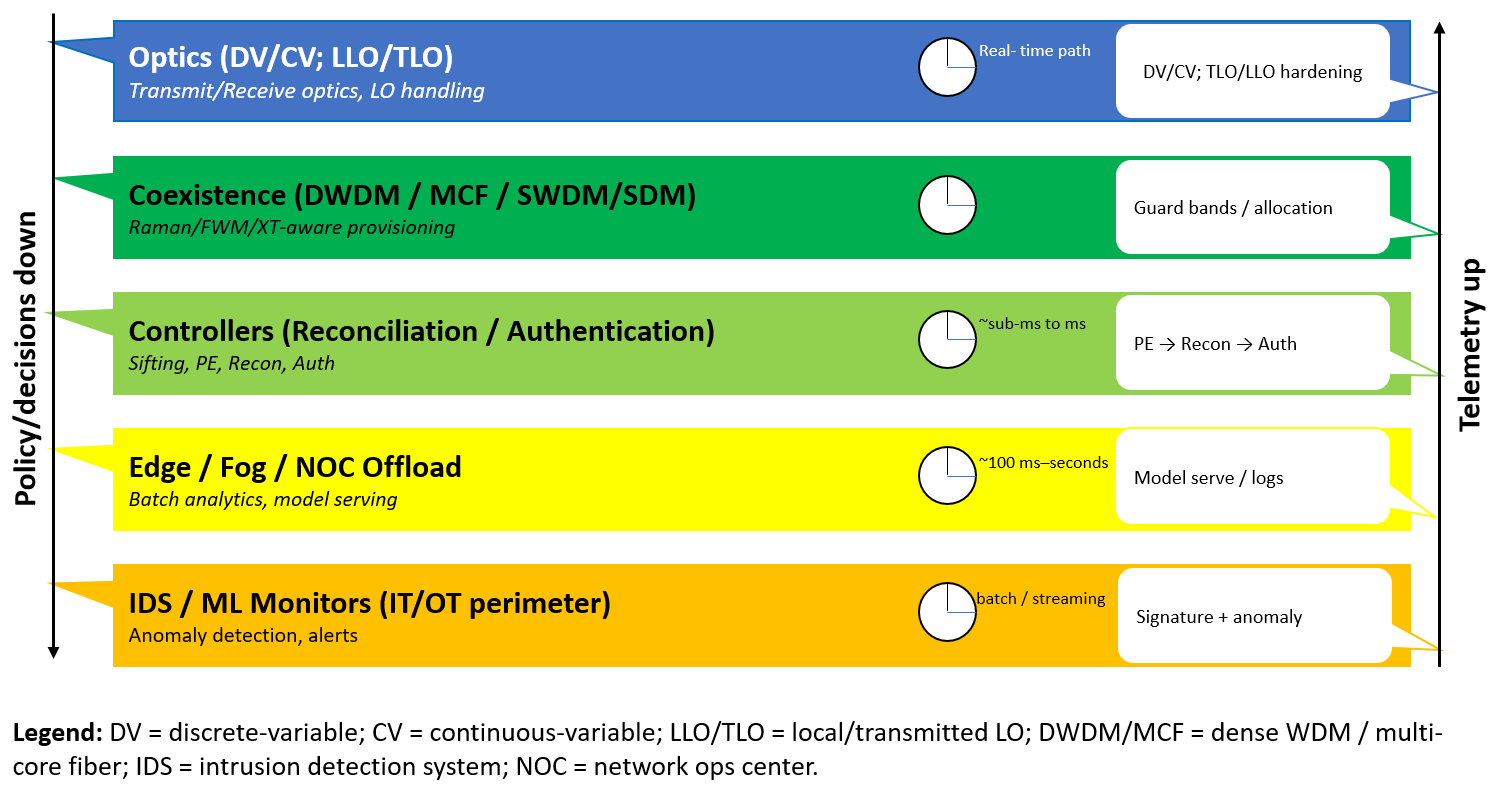}
  \caption{Deployment stack and monitoring placement. Telemetry flows upward while policy/decisions flow downward. Latency budgets indicate what can run at optics/controllers versus edge/NOC and IDS/ML.}
  \label{fig:deployment-stack}
\end{figure}

\noindent \emph{Note.} DL-based device-health monitors have demonstrated high defect-identification accuracy under EMI and environmental stress, indicating readiness for ops-side monitoring pipelines \cite{num113_Sun2025_DL_DefectID_QKD}.
The overall deployment stack and monitoring placement are illustrated in Fig.\ref{fig:deployment-stack}. This architecture integrates optical layer considerations~\cite{num11_Huang2016_LongDistanceCVQKD_ExcessNoiseControl,num12_Jouguet2013_LongDistanceCVQKD_ExperimentalDemo,num02_Hajomer2022_CVQKD60kmRLO}, coexistence provisioning~\cite{num78_Wang2022_NoisePrediction_ML_Secured_SWDM_B5G_Fronthaul,num81_Niu2019_LightGBM_NoiseSuppressing_DWDM_QKD}, hardware security constraints~\cite{num52_Zheng2021_PowerAnalysis_Integrated_CVQKD,num55_Huang2020_LaserDamage_Attenuators_QKD_PRAppl}, and edge-based resource allocation~\cite{num94_Xu2023_PrivacyPreserving_ResourceAllocation_FederatedEdge_QI_JSTSP}.

\section{Design Guidelines (Actionable)}
\label{sec:guidelines}

\noindent The synthesis in Sections~\ref{sec:problem-centric}–\ref{sec:systems} supports a small set of actionable patterns for securing DV/CV-QKD systems. We organize them around five questions a system designer actually asks: when does classical hardening suffice versus when does ML pay for itself; how should the two be combined so that failure modes do not compose; how should the ML pipeline itself be hardened; how should learned decisions be coupled to finite-key analysis; and where should monitors physically and logically sit. The guidance is grounded strictly in the surveyed corpus.

\subsection{When \emph{Classical-first} Suffices vs.\ When \emph{ML} Adds Value}
\textbf{Use classical-first} when the threat is well-characterized and has direct physical mitigations:
\begin{itemize}
  \item CV device-side channels: LO fluctuation, wavelength/calibration attacks, saturation, HD-blinding $\rightarrow$ prioritize optical filtering, power monitors, hardened calibration, detector linearity checks; consider MDI/DI shifts to reduce trust in measurement devices ~\cite{num05_Ma2013_LOFluctuationLoophole_CVQKD,num06_Huang2013_WavelengthAttack_CVQKD,num07_Ma2013_WavelengthAttack_Heterodyne_CVQKD, num08_Jouguet2013_PreventingCalibrationAttacks_LO_CVQKD,num09_Qin2016_SaturationAttack_CVQKD,num10_Qin2018_HomodyneBlinding_CVQKD,num11_Huang2016_LongDistanceCVQKD_ExcessNoiseControl,num16_Acin2007_DeviceIndependentSecurity_CollectiveAttacks,num17_Lo2012_MDIQKD_PRL,num28_Tang2014_MDIQKD_Polarization_Experimental}.
  \item DV timing/illumination vectors (time-shift, after-gate, bright-illumination/phase-remapping) $\rightarrow$ shielding/monitoring and timing randomization; MDI/DI for detector-side risk reduction \cite{num14_Lydersen2010_HackingCommercialQKD_BrightIllumination, num48_Wiechers2011_AfterGateAttack_NJP,num50_Qi2007_TimeShiftAttack_QIC,num97_Fung2007_PhaseRemappingAttack_PRA,num16_Acin2007_DeviceIndependentSecurity_CollectiveAttacks,num17_Lo2012_MDIQKD_PRL}.
  \item Coexistence impairments in SWDM/SDM/multiband optics (Raman, FWM, inter-core crosstalk) $\rightarrow$ impairment-aware channel/core planning and guard-bands~\cite{num83_Winzer2018_FiberOptic_Transmission_Networking_20Years,num79_Markowski2016_FWM_DWDM_1310nm_ApplOpt,num84_Napoli2018_TowardsMultibandOpticalSystems_PND}.
\end{itemize}
\textbf{Add ML} when adaptation or rapid screening is needed:
\begin{itemize}
  \item Parameter prediction \& stabilization/monitoring under drift (SVR/ANN/RF) in CV links and field settings~\cite{num72_Liu2017_Monitoring_CVQKD_RealEnvironment,num21_Liu2018_ML_AutomaticParameterPrediction_CVQKD,num23_Wang2019_ML_OptimalParameterPrediction_QKD,num24_Ding2020_RandomForest_ParamPrediction_QKD, num25_Su2019_BP_NN_ParameterOptimization_Atmosphere_CVQKD,num41_Liu2019_PhaseModulationStabilization_ML_QKD}.
  \item Fast detection of disturbed/attacked blocks (trees/NNs; density-based DBSCAN) and short-sample sifting screens~\cite{num30_Mao2020_ML_DetectingQuantumAttacks_CVQKD,num31_Du2022_MultiAttackDetection_NN_CVQKD, num37_Liao2022_DBSCAN_DetectingAttacks_CVQKD,num98_Wu2022_SiftingScheme_CVQKD_ShortSamples_JOSAB}.
  \item Coexistence planning at scale (noise prediction; LightGBM) to speed evaluation and placement decisions in DWDM/MCF settings~\cite{num78_Wang2022_NoisePrediction_ML_Secured_SWDM_B5G_Fronthaul,num81_Niu2019_LightGBM_NoiseSuppressing_DWDM_QKD}.
\end{itemize}

\subsection{Defense-in-Depth: Classical $\perp$ ML}
Combine orthogonal layers so that failure of one does not imply failure of all:
\begin{itemize}
  \item \textbf{Physical layer:} filtering/monitoring + architectural shifts (MDI/DI) for structural risk reduction~\cite{num11_Huang2016_LongDistanceCVQKD_ExcessNoiseControl,num12_Jouguet2013_LongDistanceCVQKD_ExperimentalDemo,num08_Jouguet2013_PreventingCalibrationAttacks_LO_CVQKD,num16_Acin2007_DeviceIndependentSecurity_CollectiveAttacks,num17_Lo2012_MDIQKD_PRL,num28_Tang2014_MDIQKD_Polarization_Experimental}.
  \item \textbf{Analytics layer:} runtime ML detectors (ANN/SVM/RF/DBSCAN) to flag anomalies that bypass physical checks~\cite{num21_Liu2018_ML_AutomaticParameterPrediction_CVQKD,num23_Wang2019_ML_OptimalParameterPrediction_QKD,num24_Ding2020_RandomForest_ParamPrediction_QKD,num30_Mao2020_ML_DetectingQuantumAttacks_CVQKD,num31_Du2022_MultiAttackDetection_NN_CVQKD, num37_Liao2022_DBSCAN_DetectingAttacks_CVQKD}.
  \item \textbf{Pre-processing defenses (CV):} apply purification (APE-GAN) or robust-manifold reconstruction (e.g., KRMNMF) ahead of detectors to mitigate adversarial perturbations~\cite{num32_Tang2023_APE_GAN_Defense_CVQKD, num58_Fu2023_KRMNMF_Defense_CVQKD}.
  \item \textbf{Process layer:} conservative estimation windows and post-selection policies to bound SKR loss if detectors trigger~\cite{num74_Furrer2012_FiniteKey_ComposableSecurity_CoherentAttacks_PRL,num75_Leverrier2017_Security_CVQKD_GaussianDeFinetti,num63_Kish2024_Mitigation_ChannelTampering_CVQKD_PRResearch}.
\end{itemize}

\subsection{Hardening ML: Data Pipelines, Adversarial Training, Certification}
\begin{itemize}
  \item \textbf{Data pipelines.} Log and version features tied to hardware and operating points (e.g., LO power, $N_0$, homodyne stats for CV; timing/count-rate for DV; coexistence proxies for SWDM/SDM) to reduce silent drift ~\cite{num05_Ma2013_LOFluctuationLoophole_CVQKD,num06_Huang2013_WavelengthAttack_CVQKD,num07_Ma2013_WavelengthAttack_Heterodyne_CVQKD, num08_Jouguet2013_PreventingCalibrationAttacks_LO_CVQKD,num09_Qin2016_SaturationAttack_CVQKD,num10_Qin2018_HomodyneBlinding_CVQKD,num11_Huang2016_LongDistanceCVQKD_ExcessNoiseControl,num12_Jouguet2013_LongDistanceCVQKD_ExperimentalDemo,num72_Liu2017_Monitoring_CVQKD_RealEnvironment,num78_Wang2022_NoisePrediction_ML_Secured_SWDM_B5G_Fronthaul,num81_Niu2019_LightGBM_NoiseSuppressing_DWDM_QKD}.
  \item \textbf{Training.} Include distribution-shift scenarios (hardware swaps, power/SNR changes, varying classical occupancy) and small-block regimes in train/validation splits~\cite{num11_Huang2016_LongDistanceCVQKD_ExcessNoiseControl,num12_Jouguet2013_LongDistanceCVQKD_ExperimentalDemo,num78_Wang2022_NoisePrediction_ML_Secured_SWDM_B5G_Fronthaul,num81_Niu2019_LightGBM_NoiseSuppressing_DWDM_QKD,num74_Furrer2012_FiniteKey_ComposableSecurity_CoherentAttacks_PRL,num75_Leverrier2017_Security_CVQKD_GaussianDeFinetti}.
  \item \textbf{Adversarial robustness.} Use adversarial training where applicable; evaluate purification (APE-GAN) and robust manifold methods (KRMNMF) as plug-in defenses; report clean vs.\ robust accuracy and induced latency~\cite{num32_Tang2023_APE_GAN_Defense_CVQKD, num58_Fu2023_KRMNMF_Defense_CVQKD,num27_Guo2023_OnePixelAttack_CVQKD,num61_Lu2020_QuantumAdversarialML_PRResearch}.
  \item \textbf{Toward certification.} While formal guarantees remain open, adopt standardized stress protocols and report worst-case performance envelopes to complement clean metrics~\cite{num61_Lu2020_QuantumAdversarialML_PRResearch,num59_Ren2022_ExperimentalQuantumAdversarialLearning, num60_Yan2023_ArtificialKeyFingerprints_CVQKD}.
\end{itemize}

\subsection{Coupling to Finite-Key Analysis}
\begin{itemize}
  \item \textbf{Decision hooks.} Expose detector confidence to key-rate calculators so that post-selection and reconciliation thresholds account for finite-size penalties (\emph{Composable} CV analyses and reductions)~\cite{num74_Furrer2012_FiniteKey_ComposableSecurity_CoherentAttacks_PRL,num75_Leverrier2017_Security_CVQKD_GaussianDeFinetti}.
  \item \textbf{Operating points.} Benchmark classical/ML defenses at matched block sizes and SNR/launch power to attribute changes in SKR and maximum distance fairly~\cite{num96_Scarani2009_Security_PracticalQKD_RMP,num11_Huang2016_LongDistanceCVQKD_ExcessNoiseControl,num12_Jouguet2013_LongDistanceCVQKD_ExperimentalDemo,num86_Kleis2019_Improving_SKR_BayesianInference_CoherentQKD}.
\end{itemize}

\subsection{Monitoring Placement \& Compute Budgets}
\begin{itemize}
  \item \textbf{Inline vs.\ side-channel taps.} Place power/LO monitors and wavelength filters inline; stream summary statistics to side analytics to avoid excess insertion loss ~\cite{num11_Huang2016_LongDistanceCVQKD_ExcessNoiseControl,num08_Jouguet2013_PreventingCalibrationAttacks_LO_CVQKD}.
  \item \textbf{Controller cycles.} Ensure ML inference meets controller loop rates; where tight, prefer trees/linear models or edge offload guided by fog/edge principles; account for embedded compute/DVFS constraints when co-locating estimation/reconciliation and ML~\cite{num53_Kim2008_PerCoreDVFS_HPCA,num44_Dastjerdi2016_FogComputing_IoT_Potential}.
  \item \textbf{Network scale.} For backbone/metro provisioning, use offline ML planning (noise prediction, allocation) with periodic re-optimization; integrate with multi-tenant and trusted/untrusted relay policies~\cite{num78_Wang2022_NoisePrediction_ML_Secured_SWDM_B5G_Fronthaul,num81_Niu2019_LightGBM_NoiseSuppressing_DWDM_QKD,num01_Cao2020_MultiTenantProvisioningQKDNet,num94_Xu2023_PrivacyPreserving_ResourceAllocation_FederatedEdge_QI_JSTSP,num95_Cao2021_HybridTrustedUntrustedRelay_QKD_Backbone_JSAC}.
\end{itemize}

\subsection{Explainability \& Safety}
\begin{itemize}
  \item \textbf{Operator-facing signals.} Prefer detectors that expose interpretable features (e.g., shifts in $N_0$, LO power, timing histograms), aiding triage and false-positive control (parallels with classical IDS operations)~\cite{num99_Alahmadi2022_CyberSecurity_Threats_SideChannel_DigitalAgriculture_Sensors,num100_Chkirbene2020_TIDCS_DynamicIDS_FeatureSelection_Access,num101_Hijazi2018_DL_IDS_IndustryNetwork_BDCSIntell,num102_Vacca2012_ComputerInformationSecurityHandbook,num103_Liao2013_IDS_ComprehensiveReview_JNCA,num104_Tama2019_TSEIDS_TwoStageClassifier_AnomalyIDS_Access,num105_GarciaTeodoro2009_AnomalyBased_IDS_Survey_CompSec,num106_Gumusbas2021_Survey_Databases_DL_Cybersecurity_IDS_IEEE_SystemsJ,num109_Sirisha2023_EfficientMLTechniques_IDS_ICSCDS,num110_Srilakshmi2022_SecureOptimizationRouting_MANETs_Access,num111_Hung2020_EnergyEfficient_CooperativeRouting_Heterogeneous_WSN_Access}.
  \item \textbf{Fail-safe defaults.} On detector uncertainty or adversarial suspicion, revert to conservative estimation/post-selection to bound SKR risk; log events for forensic analysis~\cite{num74_Furrer2012_FiniteKey_ComposableSecurity_CoherentAttacks_PRL,num75_Leverrier2017_Security_CVQKD_GaussianDeFinetti,num63_Kish2024_Mitigation_ChannelTampering_CVQKD_PRResearch}.
  \item \textbf{Artifact hygiene.} Track code/data availability and configuration (seeds, splits) to support reproducibility and safe rollbacks in production~\cite{num72_Liu2017_Monitoring_CVQKD_RealEnvironment,num98_Wu2022_SiftingScheme_CVQKD_ShortSamples_JOSAB}.
\end{itemize}

\vspace{2pt}
\noindent\textit{Deployment hint.} Start with classical hardening for known side-channels, add lightweight ML monitors for drift/attacks, and graduate to purification defenses where adversarial pressure is evidenced—always reporting SKR/distance/latency at matched operating points per the benchmarking grid in Sec.~\ref{sec:benchmarking}.


\section{Open Challenges \& Research Roadmap}
\label{sec:open}

The comparative analyses in Section~\ref{sec:problem-centric} repeatedly converged on a small set of cross-cutting gaps: matched-operating-point reporting, certified robustness, hardware-in-the-loop validation, real-time scheduling, MDI/DI vs.\ hardened-link convergence, and privacy-preserving training across operators. We treat these as the research roadmap that turns the survey's qualitative observations into a falsifiable program of work.

\subsection{Public Datasets and Benchmarks}
\textbf{Need.} Community datasets with shared feature definitions and labels for device/measurement telemetry (e.g., LO power, $N_{0}$, homodyne statistics), DV timing histograms, protocol logs (sifting, reconciliation), and coexistence traces (SWDM/SDM impairments).
\textbf{Status in corpus.} Field/lab telemetry and monitoring exist for CV-QKD and long-distance trials~\cite{num11_Huang2016_LongDistanceCVQKD_ExcessNoiseControl,num12_Jouguet2013_LongDistanceCVQKD_ExperimentalDemo,num72_Liu2017_Monitoring_CVQKD_RealEnvironment}, short-sample sifting appears in~\cite{num98_Wu2022_SiftingScheme_CVQKD_ShortSamples_JOSAB}, and coexistence traces/policies appear in~\cite{num78_Wang2022_NoisePrediction_ML_Secured_SWDM_B5G_Fronthaul,num79_Markowski2016_FWM_DWDM_1310nm_ApplOpt,num80_Kong2022_CoreWavelengthAllocation_SNS_QKD_MCF_OFC, num81_Niu2019_LightGBM_NoiseSuppressing_DWDM_QKD,num83_Winzer2018_FiberOptic_Transmission_Networking_20Years, num84_Napoli2018_TowardsMultibandOpticalSystems_PND}. 
\textbf{Action.} Release de-identified traces with split protocols, feature dictionaries, and \emph{finite-key} sweep scripts to report SKR/maximum distance at matched operating points~\cite{num74_Furrer2012_FiniteKey_ComposableSecurity_CoherentAttacks_PRL,num75_Leverrier2017_Security_CVQKD_GaussianDeFinetti}.

\subsection{Certified Robustness for QKD Attack/Anomaly Detectors}
\textbf{Need.} Beyond empirical tests, provide \emph{certified} or at least standardized robustness guarantees for ML detectors used in CV/DV settings.
\textbf{Status.} Demonstrated vulnerabilities (one-pixel, learned perturbations) and defenses (APE-GAN, robust manifold reconstruction) exist for CV-QKD~\cite{num32_Tang2023_APE_GAN_Defense_CVQKD, num58_Fu2023_KRMNMF_Defense_CVQKD,num27_Guo2023_OnePixelAttack_CVQKD}, with broader adversarial-learning context and quantum-oriented perspectives in~\cite{num61_Lu2020_QuantumAdversarialML_PRResearch,num59_Ren2022_ExperimentalQuantumAdversarialLearning, num60_Yan2023_ArtificialKeyFingerprints_CVQKD}. 
\textbf{Action.} Define adversarial test suites tailored to QKD telemetry and report clean vs.\ robust accuracy, induced latency, and \emph{worst-case} envelopes suitable for operational acceptance.

\subsection{Hardware-in-the-Loop (HIL) Training \& Evaluation}
\textbf{Need.} Train and validate CI models with the true control/measurement stack in the loop.
\textbf{Status.} Practical stabilization and monitoring pipelines are reported (phase-modulation via ML; real-environment monitoring)~\cite{num72_Liu2017_Monitoring_CVQKD_RealEnvironment,num41_Liu2019_PhaseModulationStabilization_ML_QKD}; chip/board-level concerns (power analysis, laser-damage margins) highlight cross-layer leakage~\cite{num52_Zheng2021_PowerAnalysis_Integrated_CVQKD,num55_Huang2020_LaserDamage_Attenuators_QKD_PRAppl}; controller budgets interface with embedded compute/DVFS constraints~\cite{num53_Kim2008_PerCoreDVFS_HPCA}. 
\textbf{Action.} Provide HIL harnesses that stream features from optics/electronics, enforce controller-cycle deadlines, and log SKR/latency under controlled perturbations, secure quantum federated learning and robustness for PQC/VQA pipelines~\cite{num115_Nguyen_6G_QML_AdversarialThreats_IEEE}.

\subsection{Real-Time Constraints and Scheduling}
\textbf{Need.} Guarantee inference within reconciliation/estimation control loops and sifting throughput.
\textbf{Status.} Short-sample screening and stabilization indicate tight latency envelopes~\cite{num72_Liu2017_Monitoring_CVQKD_RealEnvironment,num41_Liu2019_PhaseModulationStabilization_ML_QKD,num98_Wu2022_SiftingScheme_CVQKD_ShortSamples_JOSAB}; coexistence planning and network analytics add offline/nearline workloads~\cite{num78_Wang2022_NoisePrediction_ML_Secured_SWDM_B5G_Fronthaul,num81_Niu2019_LightGBM_NoiseSuppressing_DWDM_QKD}. 
\textbf{Action.} Publish per-block/shot latency and compute footprints alongside SKR effects; map tasks to on-device vs.\ edge/offload execution given controller cycles and power budgets~\cite{num53_Kim2008_PerCoreDVFS_HPCA,num44_Dastjerdi2016_FogComputing_IoT_Potential}.

\subsection{DI/MDI Convergence with Practical Deployments}
\textbf{Need.} Clarify when to shift to MDI/DI vs.\ harden conventional links, accounting for SKR/range/complexity and finite-size penalties.
\textbf{Status.} MDI/DI foundations are established~\cite{num16_Acin2007_DeviceIndependentSecurity_CollectiveAttacks,num17_Lo2012_MDIQKD_PRL,num28_Tang2014_MDIQKD_Polarization_Experimental}; finite-key/composable analyses for CV-QKD and security reductions inform operating regions~\cite{num74_Furrer2012_FiniteKey_ComposableSecurity_CoherentAttacks_PRL,num75_Leverrier2017_Security_CVQKD_GaussianDeFinetti}; surveys synthesize device/implementation realities \cite{num13_Xu2020_SecureQKDRealisticDevices}, \cite{num96_Scarani2009_Security_PracticalQKD_RMP}. 
\textbf{Action.} Provide head-to-head comparisons (hardened vs.\ MDI/DI) at \emph{matched} operating points and block sizes, reporting SKR, distance, and operational overheads.

\subsection{Privacy of Training Data and Federated Operations}
\textbf{Need.} Preserve confidentiality of telemetry while enabling learning across sites/vendors.
\textbf{Status.} Federated/edge learning and privacy-preserving resource allocation are explored in quantum networking contexts~\cite{num94_Xu2023_PrivacyPreserving_ResourceAllocation_FederatedEdge_QI_JSTSP}; IDS literature underscores data-sharing hurdles~\cite{num99_Alahmadi2022_CyberSecurity_Threats_SideChannel_DigitalAgriculture_Sensors,num100_Chkirbene2020_TIDCS_DynamicIDS_FeatureSelection_Access,num101_Hijazi2018_DL_IDS_IndustryNetwork_BDCSIntell,num102_Vacca2012_ComputerInformationSecurityHandbook,num103_Liao2013_IDS_ComprehensiveReview_JNCA,num104_Tama2019_TSEIDS_TwoStageClassifier_AnomalyIDS_Access,num105_GarciaTeodoro2009_AnomalyBased_IDS_Survey_CompSec,num106_Gumusbas2021_Survey_Databases_DL_Cybersecurity_IDS_IEEE_SystemsJ,num109_Sirisha2023_EfficientMLTechniques_IDS_ICSCDS,num110_Srilakshmi2022_SecureOptimizationRouting_MANETs_Access,num111_Hung2020_EnergyEfficient_CooperativeRouting_Heterogeneous_WSN_Access}. 
\textbf{Action.} Adopt federated training and privacy-aware aggregation for parameter prediction, anomaly detection, and network provisioning; document privacy/utility trade-offs.

\subsection{Standards Alignment and Reporting Discipline}
\textbf{Need.} Align CI-based defenses with commonly referenced security/operations practices (e.g., as surveyed in \cite{num13_Xu2020_SecureQKDRealisticDevices}, \cite{num96_Scarani2009_Security_PracticalQKD_RMP,granados2025quantum}) and report results in a \emph{standards-friendly} way.
\textbf{Status.} Corpus papers variably report SKR, maximum distance, and latency; explicit standards hooks are uneven.
\textbf{Action.} For each method, report: (i) Precision/Recall/F1 (or stated metrics), (ii) $\Delta$SKR and/or max distance at equal operating points, (iii) latency/compute, (iv) robustness evidence, and (v) any compliance cues (where sources explicitly mention ETSI/ITU--T)~\cite{gunasridharan2026hybrid}.

\subsection{Roadmap (Milestones)}
\textbf{Near term.} Public telemetry packs for CV/DV and coexistence; standardized adversarial/shift suites; HIL harnesses with latency counters~\cite{num72_Liu2017_Monitoring_CVQKD_RealEnvironment,num78_Wang2022_NoisePrediction_ML_Secured_SWDM_B5G_Fronthaul,num81_Niu2019_LightGBM_NoiseSuppressing_DWDM_QKD}.\\
\textbf{Mid term.} Federated training pilots with privacy accounting~\cite{num94_Xu2023_PrivacyPreserving_ResourceAllocation_FederatedEdge_QI_JSTSP}; comparative MDI/DI vs.\ hardened-link studies at matched finite-key settings~\cite{num16_Acin2007_DeviceIndependentSecurity_CollectiveAttacks,num17_Lo2012_MDIQKD_PRL,num28_Tang2014_MDIQKD_Polarization_Experimental,num74_Furrer2012_FiniteKey_ComposableSecurity_CoherentAttacks_PRL,num75_Leverrier2017_Security_CVQKD_GaussianDeFinetti}.\\
\textbf{Long term.} Robustness certification practices applicable to QKD telemetry~\cite{num32_Tang2023_APE_GAN_Defense_CVQKD, num58_Fu2023_KRMNMF_Defense_CVQKD,num27_Guo2023_OnePixelAttack_CVQKD,num61_Lu2020_QuantumAdversarialML_PRResearch}; integrated deployment guides that tie CI components to end-to-end SKR/distance and operational SLOs \cite{num96_Scarani2009_Security_PracticalQKD_RMP,num13_Xu2020_SecureQKDRealisticDevices,num11_Huang2016_LongDistanceCVQKD_ExcessNoiseControl,num12_Jouguet2013_LongDistanceCVQKD_ExperimentalDemo}.


\section{Conclusion}
\label{sec:conclusion}
This survey took a problem-driven view of DV/CV-QKD, contrasting classical countermeasures with ML-enabled techniques across five layers—device/measurement, channel, protocol/finite-key, ML, and network/coexistence—and analyzing each problem class under a single \emph{problem $\rightarrow$ classical $\rightarrow$ ML $\rightarrow$ comparison $\rightarrow$ gaps} template. The recurring finding across P1–P9 is that the gap between provable and practical security is no longer best characterized as a gap in security models; it is a gap in \emph{measurement discipline}. The corpus contains strong nominal numbers—DBSCAN attack detectors at F1 $=0.998$, decision-tree channel-tampering detectors at 100\%/91.26\% across noise regimes, post-selection SKR gains of $+0.024$ bits/pulse, robust adversarial accuracies of 71.6–79.5\%, and LightGBM-style coexistence planners cutting evaluation time by up to 98.8\%—but these numbers are reported at heterogeneous operating points and rarely paired with the $\Delta$SKR, distance, and controller-cycle latency they ultimately compete in.

Classical-first mitigations (optical filtering, power monitoring, calibration hardening, MDI/DI, impairment-aware planning) remain the auditable backbone of any deployable QKD link, while ML earns its place where adaptation and rapid screening matter most: drift-robust parameter prediction, multi-attack and defect classification, adversarial purification, and large-scale coexistence planning. ML also imports its own attack surface—distribution shift, adversarial perturbations, and accountability of learned components—that must be explicitly managed rather than treated as residual risk.

We therefore proposed a benchmarking template—shared datasets (fiber/free-space; lab/field), stress protocols (adversarial, distribution shift, small-block finite-key), and consistent reporting (P/R/F1, $\Delta$SKR and maximum distance at matched operating points, latency/compute/energy, robustness evidence)—together with design guidance built around five principles: defense-in-depth that composes classical and ML \emph{orthogonally}, robust and versioned data pipelines, adversarial training or inline purification where adversarial pressure is evidenced, explicit coupling of detector decisions to finite-key analysis, and careful monitor placement within controller cycles. The priority next steps that follow directly from these gaps are public telemetry and stress suites, hardware-in-the-loop evaluation, head-to-head MDI/DI versus hardened-link studies at matched finite-key settings, privacy-preserving federated training, and standards-aware reporting—each chosen so that completing it would let the corpus' existing numbers be combined into auditable, real-world QKD deployment recommendations.





\bibliographystyle{IEEEtran}
\bibliography{references}

\vfill

\end{document}